\documentclass[twocolumn]{svjour3} 
\smartqed 
\usepackage[tight]{subfigure}
\usepackage{balance}
\usepackage{comment}
\usepackage[justification=centering,small]{caption}

\usepackage{algpseudocode}
\usepackage{algorithm}
\usepackage{mathtools}
\DeclarePairedDelimiter\ceil{\lceil}{\rceil}

\begin{document}

\title{Towards Coordinated Bandwidth Adaptations for Hundred-Scale 3D Tele-Immersive Systems}

\author{Mohammad Hosseini       \and
        Gregorij Kurillo       \and
        Seyed Rasoul Etesami      \and
        Jiang Yu       
}

\institute{Mohammad Hosseini\at
              Department of Computer Science\\
              University of Illinois at Urbana-Champaign\\
              \email{shossen2@illinois.edu}       
           \and
           Gregorij Kurillo \at
              Electrical Engineering and Computer Sciences Department\\
              University of California at Berkeley\\
              \email{gregorij@eecs.berkeley.edu}          
           \and
           Seyed Rasoul Etesami \at
              Department of Computer and Electrical Engineering\\
              University of Illinois at Urbana-Champaign\\
              \email{etesami1@illinois.edu}           
                         \and
           Jiang Yu \at
              Department of Computer Science\\
              University of Illinois at Urbana-Champaign\\
              \email{jy1989@illinois.edu}           
              }

\date{Received: July 30, 2015}

\maketitle

\begin{abstract}
3D tele-immersion improves the state of collaboration among geographically distributed participan-ts. Unlike the traditional 2D videos, a 3D tele-immersive system employs multiple 3D cameras based in each physical site to cover a much larger field of view, generating a very large amount of stream data. One of the major challenges is how to efficiently transmit these bulky 3D streaming data to bandwidth-constrained sites. In this paper, we study an adaptive Human Visual System (HVS) -compliant bandwidth management framework for efficient delivery of hundred-scale streams produced from distributed 3D tele-immersive sites to a receiver site with limited bandwidth budget. Our adaptation framework exploits the semantics link of HVS with multiple 3D streams in the 3D tele-immersive environment. We developed TELEVIS, a visual simulation tool to showcase a HVS-aware tele-immersive system for realistic cases. Our evaluation results show that the proposed adaptation can improve the total quality per unit of bandwidth used to deliver streams in 3D tele-immersive systems.

\end{abstract}

\section{Introduction}
\label{sec:intro}
Nowadays with the growing demands of networked applications along with the advancements in networking technologies with support of high bandwidth, while traditional applications such as email, web browsing, music, and video are still popular, there has been a significant rise in the new 3D media such as 3D gaming, 3D videos, 3D virtual environments, and 3D tele-immersive systems. Coming along with Internet 2, 3D tele-immersion technology takes place when individuals at geographically separated sites try to collaborate in a 3D shared virtual environment. Users of an immersive site, or a performer, produce 3D streams by a 3D camera array surrounding their physical space. A 3D model of the performer will then be constructed from the streams and placed on the virtual stage to interact with other users. Thus, the performers can be physically dispersed. The applications are not only used for entertainment, video games and interactive storytelling \cite{ImmersiveApp1}, but also used for other purposes such as biology and tele-medicine for remote health-caring \cite{ImmersiveApp2,ImmersiveApp4}, art \cite{ImmersiveApp3}, as well as military and education \cite{ImmersiveApp1}. Major ongoing research is carried out covering different aspects to deliver 3D immersive experience in real time \cite{3dti1,3dti2,3dti3,3dti4}.

Despite the promising nature of 3D tele-immersion systems, many challenges still remain to make this technology more feasible and available to everyday use. Whi-le we are already witnessing the extreme growth of many live video broadcasting services such as Youtube \cite{youtube}, Livestream \cite{livestream}, and CoolStreaming \cite{coolstreaming}, existing applications of 3D Tele-immersion systems are still restricted with a small number of users. Unfortunately due to the resource limitations such as network bandwidth, one of the major challenges to increase the number of performers and 3D cameras is how to efficiently transmit the bulky 3D stream data to bandwidth-const-rained distributed sites. Unlike the traditional 2D systems such as video conferencing, 3D tele-immersive systems use an array of 3D cameras which in overall leads to a bulky amount of data needed to be transmitted over the networks. Even for a basic setting, the data rate requirement of a single 3D stream may reach up to 100 Mbps and in case of ten 3D cameras in a physical site, the total bandwidth requirement could surpass Gbps level~\cite{yang}. Thus there must be a balance between the requirements of 3D streaming and the architecture of Internet2 backbone. One of the challenges to achieving this balance is to meet the demands for a viewing preference given the limited bandwidth without much negative impact on the user's experience. However our work is motivated by the idea of 3D Tele-immersive amphitheaters with hundred-scale users \cite{shannon} and the bandwidth issues that such rich multimedia system arises, a semantic relation between HVS requirements and multi-ranked stream priorities has not been fully developed for the purpose of the bandwidth management and high performance tele-immersion protocols over Internet2 networks in the previous work. Hence, we propose to utilize this semantic relation in our novel stream adaptation framework, in order to address the concerns and challenges that are neglected by previous research.

In this paper, we study an adaptive HVS-compliant stream management framework for efficient delivery of hundred-scale streams to distributed sites with limited bandwidth budget over networks. Our adaptation framework relies on the concepts of HVS-compliant Field of View (FOV) which exploits the semantics relation of multi-priority 3D streams in the 3D tele-immersive environment, with the main goal of maximizing the total quality of the streamed data to satisfy a bandwidth budget specific to the receiver site. Because of the lack of a fine-grained HVS-compliant prioritized component, many 3D tele-immersion systems extensively experience inefficient resource usage due to the fact that they manage all streams in the same way, with no distinguished importance. Our approach is to selectively choose the proper frame rate, i.e. Frame Per Second (FPS) of each 3D camera given their stream priorities based on FOV of receiver's rendered virtual viewpoint, by limiting the total amount of streamed data to a viewer site while satisfying the bandwidth budget. Our evaluation results show that using our adaptation framework significantly improves the quality per unit of bandwidth consumed to transmit the stream data while it adapts streams according to the users' view preference. In summary, the contributions of this paper include the following: 

\begin{itemize}
\item Feasibility study of a coordinated adaptive stream management system, which enhances the utilization of limited bandwidth issues by incorporating users' view requirements and adaptation.
\item Efficient bandwidth management in regards to the hundred-scale prioritized streams, which fine-grains the differentiation of streams and improves the resulting qualities. The computational cost induced by our adaptation mechanism is trivial that makes it suitable for scalability requirements of 3D tele-immersive systems with high number of streams.
\item Development of TELEVIS, a visual simulation tool with a configurable GUI to showcase our HVS-comp-liant adaptations for tele-immersive systems.
\end{itemize}

It should be noted that this paper is an extended study of our previous conference version \cite{mmve15}, which is extended by 70\% relative to the prior study. The new contribution includes a) development of TELEVIS visual simulation tool, which models the interaction of participants using facing and clustering algorithms, b) comparison of the relative execution time overhead of our proposed algorithm against BLMCK algorithm \cite{blmck}, tested for large number of streams, and c) the discussion of runtime overhead for the main proposed heuristic in Section 3.4. In addition, our extended paper discusses more details, including an extended introduction and more thorough discussion of related work.

The paper is organized as follows: in Section \ref{sec:background}, we cover a wide area of related work, and we discuss how our system is related to them. Section \ref{sec:methodology} explains our methodology for adaptive stream management system, including the problem definition and the stream prioritization. Our evaluation results are presented in Section \ref{sec:evaluation}, while in Section \ref{sec:conclusion} we conclude the paper and briefly discuss possible improvements for future work.

\section{Related work}
\label{sec:background}
In this section we present different concepts and categories of work relevant to our proposed framework.
\subsection{Human Visual System (HVS)}
\begin{figure}[!t]
\centering
\includegraphics[width=2.2in]{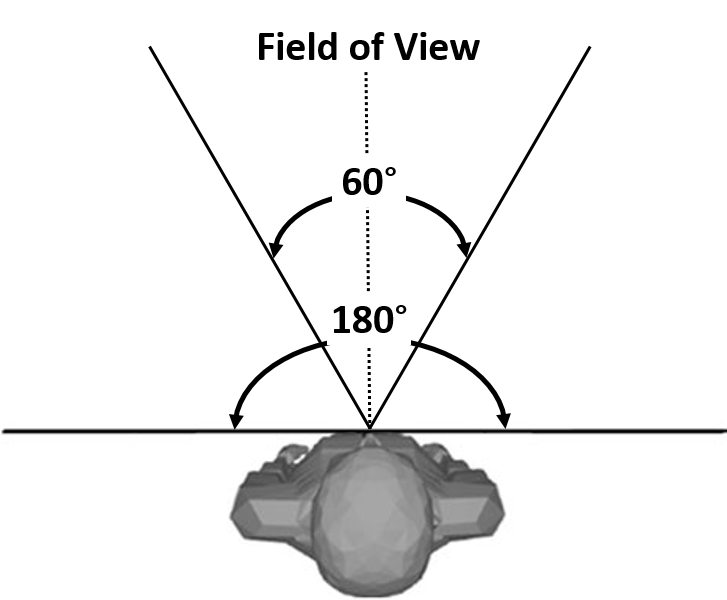}
\caption{Forward-facing horizontal FOV in HVS}
\label{fov}
\end{figure}

There are many parameters that naturally most of them being the attributes that constitute the human visual system, which ultimately determine the quality of the acquired raw visual input. One of the important visual parameter is the Field of View (FOV), the angle subtended by the viewing surface from a given observer location. Humans have an FOV of about 180 degrees horizontal, allowing them to have an almost 180-degree forward-facing horizontal FOV, while the central focus angle of view is approximately 40-60 degrees nasally, directly in front of the eyes and is actually what most influences human perception of a scene, as shown in Figure \ref{fov}. This is an important factor in design of various display devices such as monitors to provide a wide FOV to the users \cite{hvs1}, \cite{hvs2}, \cite{hvs3}. We aim to leverage these unique HVS features towards design of our prioritized streaming management system which well suits the requirements of 3D tele-immersion systems. In order to enable that, we are making two assumptions for our design. Firstly, we are assuming that the central vision is fixed to the head orientation, and is directly applied to Head-Mounted Displays (HMDs). Iris recognition algorithms can be considered in this regard for detection of gaze direction~\cite{iris1}. In order to apply our design to a general display, we need to take into account the dependency of FOV on direction of the gaze and distance from the screen~\cite{msr,realistic}. The second assumption is related to the zoom factor. We are assuming that in the virtual reality, the scale is 1:1 where the users probably will not zoom in on the scene. However, our design allows to accommodate FOV changes for zooming scenarios. Also, it should be noted that we are considering observations from first person perspective which is different from most of the prior work with for example remote dancers \cite{Dance2,Dance1}.

\subsection{Dynamic Adaptive Streaming}
Adaptive streaming in general is a process whereby the quality of a multimedia stream is altered while it is being sent from a server to a client. This topic has been investigated since late 90's during the days of multicast video conferencing. As an example, Elan Amir \textit{et al.} proposed a protocol called SCUBA \cite{amir} that enables media sources to intelligently account for receiver interest in their rate-adjustment algorithms. Generally, quality adaptation is controlled by decision modules on either the \textit{client} or the \textit{server}, and may be the result of adjusting various network or device metrics~\cite{dash2}. For example, with a decrease in network throughput, adaptation to a lower video bitrate may reduce video packet loss and improve the user's experience. Dynamic Adaptive Streaming over HTTP (DASH) in specific, also known as MPEG-DASH, provides adaptive bitrate streaming where a video file is partitioned into several segments and delivered to a client using HTTP. MPEG-DASH is the first adaptive bit-rate HTTP-based streaming solution which is currently an international standard~\cite{dash1}.

\subsection{Coordinated Internet Video Control Plane}
Video traffic already represents a significant fraction of today's Internet traffic and is projected to exceed 90\% by the end of 2014 \cite{video}. Liu \textit{et al.} in their study \cite{liu1} made a case for a video control plane combined with video bitrate selection/adaptation using large-scale measurements gathered from over 50 million users. In their work it is assumed that a video content provider such as YouTube or Hulu runs a control plane to monitor and improve the video experience for its customers. The control plane can use a global view of network and CDN performance to dynamically assign clients a suitable choice of bitrate that optimizes the video delivery. Beyond the performance benefits, such a control plane also offers content providers more flexibility in instrumenting fine-grained policies; e.g., providing higher quality service to premium customers under load, ensuring that certain types of content is only accessible with specific geographical regions, or taking into account the cost-performance tradeoffs that different CDNs have to offer \cite{liu2}. We adopt the same concept of coordinated control plane towards design of a coordinated 3D stream management system.

\subsection{Prioritized Multimedia Streaming}
Generally the priority of multimedia contents originated from various sources can have different importance given the settings and the multimedia context. Hosseini \textit{et al.} \cite{hosseini1}, \cite{hosseini2}, \cite{hosseini3} adopted priority-based approaches towards the study of efficiently transmitting, rendering, and displaying bulky 3D gaming information to power-limited devices given the importance of different 3D objects in the gaming context. Also in the context of telepresense, DeVincenzi \textit{et al.} \cite{priority1}, applied the concept on a smart multi-user teleconference room equipped with a static camera capturing the whole room. In the beginning of speaking, the system sends notifications to the video encoder component to assign higher video resolution to the speaker's location in the scene.

\subsection{3D Multimedia Adaptation}
The idea of 3D multimedia adaptation has been used in several multimedia streaming contexts including tele-immersion systems. While many of the available systems use \textit{image-based} rendering and fast frame compression, these schemes have been a bottleneck in deployment of 3D tele-immersive systems due to high resource demands in terms of both memory and bandwidth \cite{3dmm5,3dmm2}. Kum \textit{et al.} proposed a real-time compression scheme \cite{3dmm6} for a multi-view plus depth-based 3D environments to solve the 3D reconstruction issues of 3D cameras. Their system clusters streams together and predicts all streams in a cluster from a single stream. Wurmlin \textit{et al.} implemented a 3D video acquisition environment for virtual design and collaboration \cite{3dmm3}, that processes inputs from multiple cameras, merges them into a single 3D video stream which is then transmitted to a remote site. In their system a dynamic camera control system adjusts the 3D video process for enhanced performance. Also Yang \textit{et al.} \cite{yang} proposed a hierarchical system for 3D compression and adaptation of 3D content to achieve higher rendering quality. However, generally the concern of FPS-based bandwidth adaptation in relation to the HVS requirements of participants, multi-ranked prioritization of streams, and the corresponding quality trade-offs has remained unaddressed in these related works.

~\newline
In this paper, we build on top of concepts from all the aforementioned related work, and design an adaptive coordinated framework for efficient delivery of streams to 3D tele-immersive sites with a limited bandwidth budget in regards to the users' HVS.

\section{Proposed Framework}
\label{sec:methodology}

\begin{figure*}[!t]
\centering

\includegraphics[width=\textwidth]{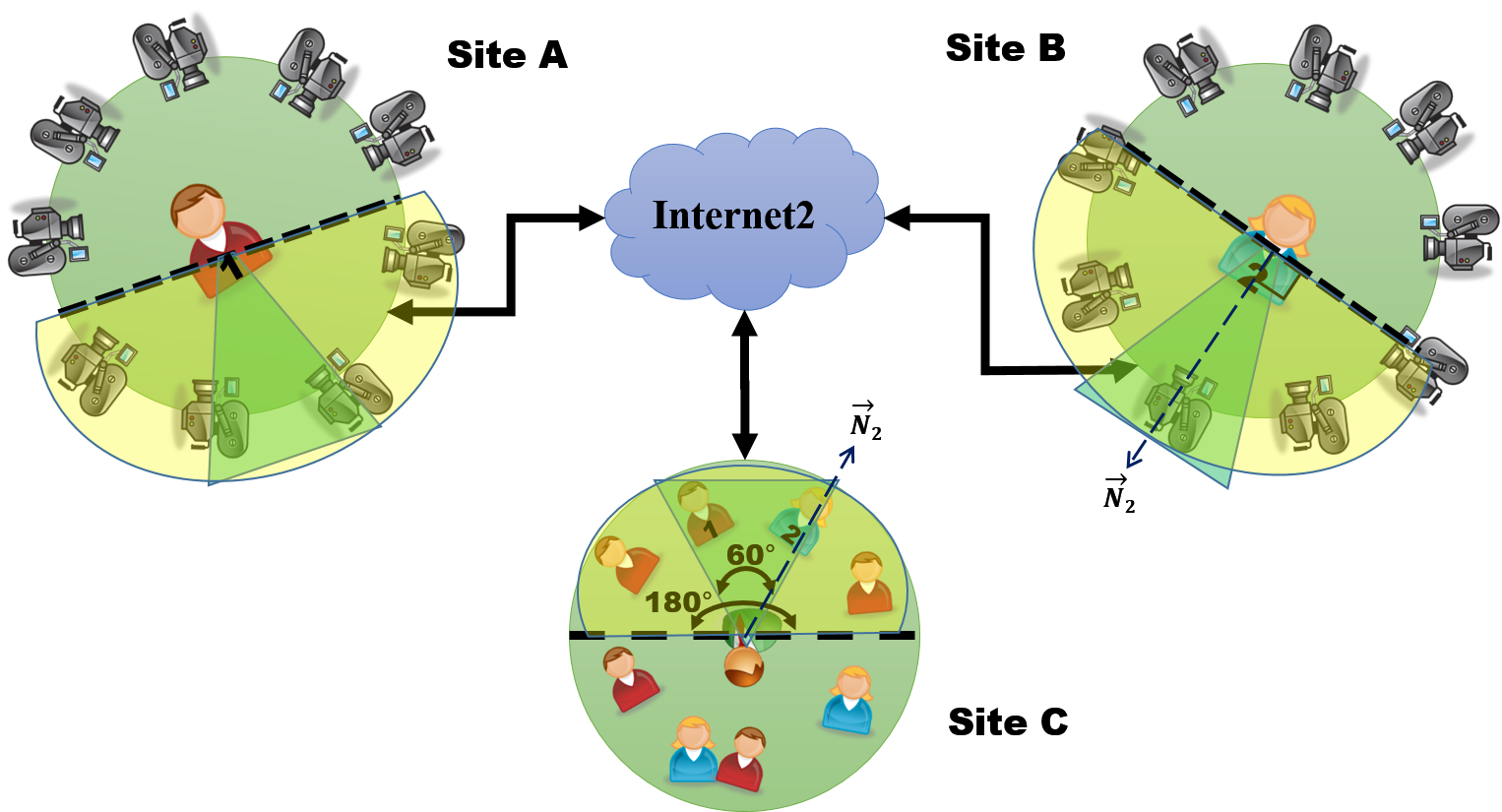}
\vspace{.2cm}
\caption{An example visual view of our HVS-compliant prioritized streaming model}
\label{scheme}
\end{figure*}

\begin{figure}[!hp]
\centering
\includegraphics[width=\columnwidth]{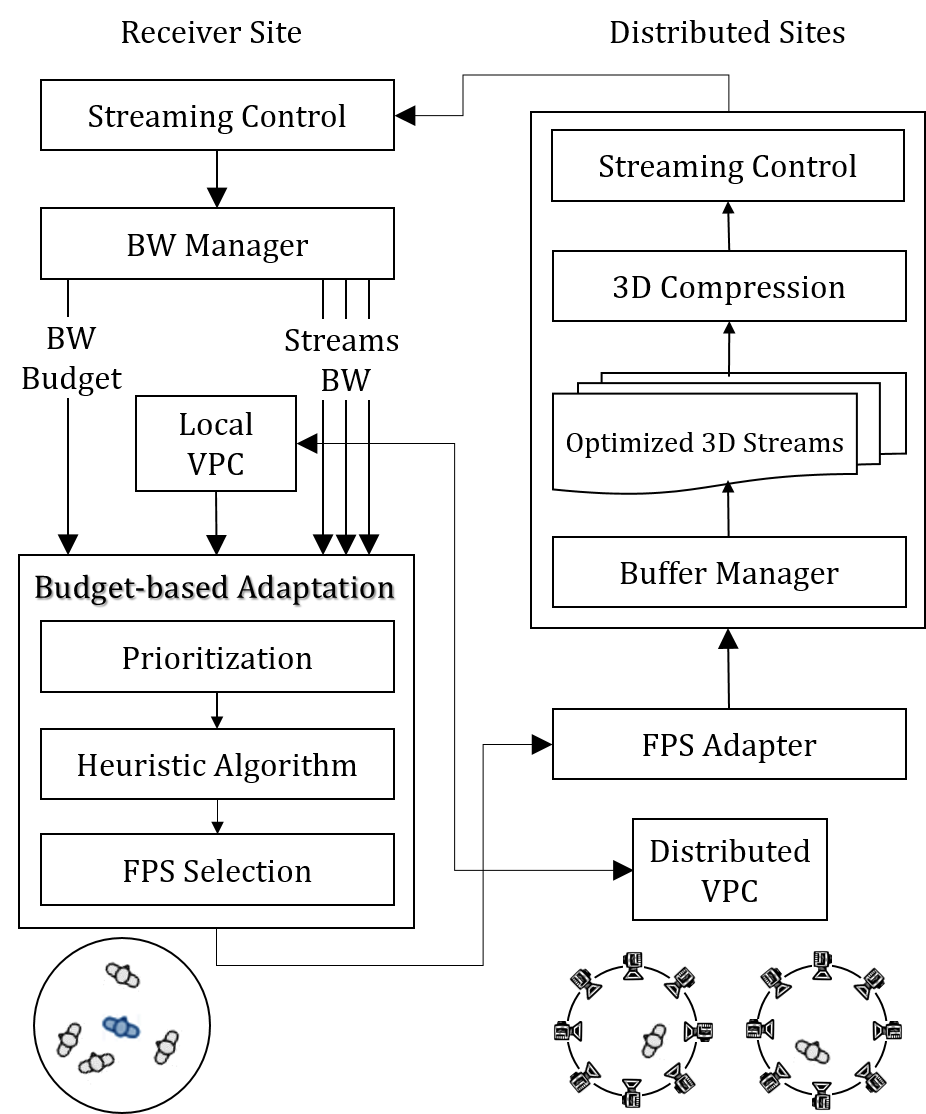}
\vspace{.1cm}
\caption{Outline of the proposed framework. Streams generated at multiple distributed (remote) sites are transmitted to the receiver (local) site.}
\label{framework}
\end{figure}

Our design allows for efficient transmission of streams to receiver sites in regards to the viewer's FOV while satisfying a bandwidth budget. Figure \ref{scheme} shows a visual view of our HVS-compliant prioritized design modeled for three sites, with site A and site B being performers and site C being a passive viewer, showing potential virtual locations of other performers. This circular configuration is considered with full visibility in the overlapping space. As is shown in the figure, the design of our bandwidth adaptation framework revolves around the concept of HVS requirements and a two-level stream priority. The priorities are considered in regards to both the sites corresponding to each distributed active participant, and also the cameras within each site. The first-level priority is associated with the distributed participants in the virtual space depends on each viewer's angular FOV on the receiver site, and the second-level priority depends on the angular coverage of cameras within each participant's physical site. Under these prioritization, the framework will differentiate among streams according to their contribution to the current FOV and select suitable adaptation per stream for dynamic bandwidth management and FPS adjustment. We adopt two View-aware Priority Calculator (VPC) modules to provide priority contributions:

The local VPC at the receiver site is responsible for the first-level priority. Given the fact that the user is only interested in one particular participants, the local VPC captures the user's FOV changes to calculate at any time, which participants are covered within the main FOV (within front horizontal $60^{\circ}$), and which ones are covered by the wide FOV of the user (within $180^{\circ}$). Apparently the framework ignores the participants who are not in the wide FOV of the users as they are not visible.

The distributed VPC at each distributed participant site controls the second-level priority. It calculates $N_i$, the corresponding normal vector perpendicular to each participant $i$ presented in the virtual space in respect to the receiver's wide FOV orientation, which is then used to calculate the priority of cameras at the participant's physical site. Similar to the local VPC, the distributed VPC computes the coverage of cameras within front $60^{\circ}$ main angle and wide $180^{\circ}$ angle, both bisected by $N_i$. Note that $N_i$ is not necessarily in the center of the participant's view cone, and is actually constructed in respect to the view dependency of the receiver towards each participant. Distributed VPC likewise disregards the cameras located on the exterior wide $180^{\circ}$ as they contribute to the streams which is not visible to the receiver.

Figure \ref{framework} shows a detailed overview of different modules and processes in our framework. As can be seen in the figure, the system consists of modules both at each distributed site, and the receiver site visualizing potential virtual locations of distributed users. As a part of the receiver site, the bandwidth manager module determines the bandwidth budget and the bandwidth contribution of each specific stream, which is given as input to the main adaptation module.

The first-level priority and second-level priority contribute to a global priority. The bandwidth budget, the streams bandwidth feed, along with the global priority are then used to provide adaptation by the heuristic algorithm that we will describe in detail in the next subsection. Under the budget-based adaptations, proper FPS of each stream for all distributed site is selectively calculated and is sent to the FPS Adapter module at each distributed site. The FPS Adapter then applies the specified FPS to the cameras' buffer manager. The resulting 3D streams are then compressed, serialized, and streamed to the receiver through the gateway to be finally decompressed, rendered and visualized in parallel the bandwidth manager monitors used bandwidth and determines the new bandwidth budget in case of variations and the bandwidth contribution of each stream. Receiving optimal and efficient stream bitrate not only reduces the network bandwidth requirement and variations, but also less exhausts other limited resources (such as memory and processor units) to buffer, process, and render less amount of data.

\subsection{Problem Modeling}
For stream selection and transmission in 3D tele-immer-sive systems, one of the most significant factor in bandwidth usage is the FPS which corresponds to the amount of data produced by the 3D cameras. As discussed in Section \ref{sec:intro}, we suppose that the receiver site specifies a bandwidth budget. If the total bandwidth requirement of the streams does not exceed the bandwidth budget, then all of them can be received. If the bandwidth budget \textit{W} is insufficient to receive all the streams, then the total bandwidth must be reduced. Every 3D camera produces a specific stream bandwidth, and we must decide how to adopt a proper FPS for each camera given the HVS requirements so to be able to receive the streams within the budget \textit{W}.

One approach to reducing the total stream bandwidth is to transmit a subset of the required streams. This selection scheme is the well-known binary Knapsack optimization problem. The binary Knapsack problem is NP-hard but efficient approximation algorithms can be utilized (fully polynomial approximation schem-es), so this approach is computationally feasible. However, this method enables to only select a \textit{subset} of the streams, which is not desired since the user intends to receive \textit{all} the necessary streams. Our proposed algorithms selects all necessary streams, but with different bandwidth requirements according to their priorities. This is a multiple-choice knapsack problem in which the items (streams) are organized into groups corresponding to the objects. Each group contains the full-bandwidth stream corresponding to an object and lower-bandwidth versions of the same streams obtained by applying FPS adaptation. If computationally feasible, segmentation-based video coding techniques can also be employed to further reduce the bitrate of streams in case of lower bandwidth demands \cite{seg2,seg1}. Overall, the goal is to choose one stream from each group within a bandwidth budget $W$.

\subsection{Stream Adaptation}
\label{sub:adapt}
If streams need high bandwidth, but are less important for the user's view, then the gateway servers on both the receiver site and the distributed sites are wasting resources transferring, and rendering them. Thus the first step in our method is to establish the global priority of each camera stream within a specific 3D tele-immersion scene. Given the first-level and second-level priorities by the VPC modules, the prioritization module classifies the streams into four classes of pairs (first-level priority, second-level priority), namely $C_{11}$, $C_{12}$, $C_{21}$, and $C_{22}$, with $C_{11}$ representing the ``low", $C_{12}$ and $C_{21}$ the ``medium", and $C_{22}$ the ``high" priority classes. 

Our problem is to receive streams in a way that maximizes the total quality of the streams within a user-specified bandwidth budget. The bandwidth requirement of a stream can be reduced by reducing its FPS, but this also reduces the quality. To take this into account, we set a user-defined maximum level of reduction $R_{max}$ which is the maximum reduction in FPS that is acceptable to the user. We assume that the quality of a stream is a function of its bandwidth requirement (with maximum quality corresponding to minimum FPS reduction) and its global priority. Without loss of generality and to enable a robust quantitative modeling in the simulation environment, we use the product of bandwidth and global priority as the simplest function to the measure of quality in this study, similar to work in \cite{mmve15}, \cite{hosseini1}, and \cite{hosseini2}. For example, a stream $\tau_i$ of original bandwidth $s_{\tau_i}$ with FPS scaling factor $r_i$ (therefore, the resulted frame rates are factorials of the camera frame rate require no interpolation processing at the sender side), and global priority of $p_{\tau_i}$ contributes to quality $s_{\tau_i}\times p_{\tau_i}\times r_i$. Our goal is to apply stream bandwidth reduction in a way that maximizes the total quality subject to the constraints of the bandwidth budget $W$ and maximum reduction $R_{max}$.

\subsection{Proposed Solution}
\label{sub:sol}
\begin{algorithm}[!t]
\begin{algorithmic}
\State $\mathcal{T}$: prioritized list of streams sorted from smallest to largest
\State $\tau_i$: stream with full-bandwidth $s_{\tau_i}$
\State $x_i$: adapted stream with bandwidth $s_{x_i}$
\State $R_{max} = 1/c^k$: maximum reduction of bandwidth
\State Calculate $W_0 = \sum s_{\tau_i}\times R_{max}$ \%comment: minimum bandwidth requirement
\State $\forall \tau_i \in \mathcal{T}: s_{x_i} \gets s_{\tau_i}\times R_{max}$ \%comment: apply $R_{max}$ to all $\tau_i$'s.

\While {$s_{\tau_i}\times (1-R_{max}) \leq W_{i-1}$} \\ 
\%comment: i=1 initially.
\State $s_{x_i} \gets s_{\tau_i}$
\State $W_i \gets W_{i-1} - s_{\tau_i}\times (1-R_{max})$
\State{ $i \gets i+1$~~\%comment: adapt next stream}
\EndWhile \\
\%comment above loop repeats until some streams $\tau_{\ell}$ cannot be received at full bandwidth within the remaining budget $W_{\ell-1}$.
\State $\ell \gets i$ \%comment: resulting from above loop.
\State{ Find minimum $r_{l}=1/c^l$ such that \\$s_{\tau_{\ell}}\times r_{l}\leq W_{\ell-1}+ (s_{\tau_{\ell}}\times R_{\max})$ ~~\%comment: determines the maximum bandwidth at which $\tau_{\ell}$ can be received by calculating the minimum bandwidth reduction $r_{\ell}$.}
\State{ $s_{x_{\ell}} \gets s_{\tau_{\ell}}\times r_{l}$ \%comment: adapt $\tau_{\ell}$ and calculate $s_{x_{\ell}}$}
\end{algorithmic}
 \caption*{Algorithm 1: Compromise}
\end{algorithm}
There are $n$ streams $\mathcal{T}=\{\tau_1,\tau_2,\ldots,\tau_n\}$, and each $\tau_h \in \mathcal{T}$ has an original bandwidth requirement of $s_{\tau_h}$, and a global priority or importance $p_{\tau_h}$. The quality of stream $\tau_h$ is $q_{\tau_h} = p_{\tau_h}\times s_{\tau_h}$. The duration of the 3D tele-immersive session is $T$, and the bandwidth budget limits the total bandwidth of the streams that can be received at the receiver site to $W$.

Let $X= \{x_1,x_2,\ldots,x_n\}$, be the set of streams that are received at the receiver site. Each $x_i \in X$ corresponds to an original stream $\tau_i \in \mathcal{T}$. Each $x_i$ has an arrival time $a_{x_i}$ when it becomes visible, a departure time $d_{x_i}$ after which the stream can be ignored, and global priority $p_{x_i} = p_{\tau_i}$.

The bandwidth requirement of a specific stream $x_i \in X$ is the original bandwidth requirement of the stream $\tau_i \in \mathcal{T}$ scaled by a factor of $c^{r_i}$ for some $0 \leq r_i \leq k$ and constant $c$ where $R_{max}=`/c^k$ is specified by the user as the \textit{maximum reduction of bandwidth} that can be tolerated. Our choice of a scaling factor of $c^{r_i}$ was motivated by the systems based on Distributed Hash Table such as Chord \cite{chord} in which the distance between a node and its fingers increases exponentially. So, the bandwidth requirement of $x_i$ is $s_{x_i}=\frac{s_{\tau_i}}{c^{r_i}}$, and the quality of the stream $x_i$ is $q_{x_i}=p_{x_i}\times s_{x_i}= \frac{p_{\tau_i}\times s_{\tau_i}}{c^{r_i}}$.

\subsection{Algorithms}
\begin{algorithm}[!t]
\begin{algorithmic}
\State $\mathcal{T}$: prioritized list of streams sorted from smallest to largest
\State $\tau_i$: stream with full-bandwidth $s_{\tau_i}$
\State $x_i$: adapted stream with bandwidth $s_{x_i}$
\State $R_{max} = 1/c^k$: maximum reduction of bandwidth
\While {$\sum s_{x_i} \leq W$}
\State Find maximum $j<k$ such that $r_{i}=1/c^j$ and \\$s_{\tau_{i}}\times r_{i}\leq W_{i-1}+ (s_{\tau_{i}}\times R_{\max})$ ~~\%comment: determine minimum bandwidth reduction $r_{i} \geq R_{\max}$
\State $s_{x_{i}} \gets s_{\tau_{i}}\times r_{i}$ ~~\%comment: adapt $\tau_{i}$ and calculate $s_{x_{i}}$
\State $i \gets (i+1)~\%~n$~~\%comment: adapt next stream; if the end, start from the beginning
\EndWhile
\end{algorithmic}
 \caption*{Algorithm 2: Round-Robin}
\end{algorithm}

Let $S$ be the total bandwidth of all streams, and $W$ be the bandwidth budget. The maximum reduction of bandwidth that the user will tolerate is $R_{max} = 1/c^k$. Let $C_{11}$ be the class of streams with the smallest global priority, and similarly for $C_{12}$, $C_{21}$, and $C_{22}$. 

For each stream $\tau_i$ in the list, we calculate $q_i$ as described in the previous subsection. This is the contribution that $\tau_i$ would make to the average quality of the system if it were received at full bandwidth. We then calculate $W_{\min} = S\times R_{\max}$ which is the minimum bandwidth budget that is needed to receive all streams. If $W_{\min}>W$, then the problem cannot be solved. In the following, assume that $W_{\min} \leq W$ so the unused bandwidth budget is $W_0=W-W_{\min}$.

\begin{algorithm}[!t]
\begin{algorithmic}
\State $\mathcal{T}$: prioritized list of streams sorted from smallest to largest
\State $\tau_i$: stream with full-bandwidth $s_{\tau_i}$
\State $x_i$: adapted stream with bandwidth $s_{x_i}$
\State $R_{max} = 1/c^k$: maximum reduction of bandwidth
\While {$\sum s_{x_i} \leq W$}
\Repeat \State{ Find maximum $j<k$ such that $r_{i}=1/c^j$ and \\$s_{\tau_{i}}\times r_{i}\geq W_{i-1}+ (s_{\tau_{i}}\times R_{\max})$ ~~\%comment: determine minimum bandwidth reduction $r_{i} \geq R_{\max}$}
\State{ $s_{x_{i}} \gets s_{\tau_{i}}\times r_{i}$ \%comment: adapt $\tau_{i}$ and calculate $s_{x_{i}}$}
\Until{$j=k$} ~~\%comment: $r_{i}=R_{max}$
\State{ $i \gets i+1$~~\%comment: adapt next stream}
\EndWhile
\end{algorithmic}
 \caption*{Algorithm 3: Aggressive}
\end{algorithm}

To determine the bandwidth reduction for each stre-am, this algorithm (namely \textit{Compromise}) sorts the prioritized list of incoming streams by the global priority from the largest to the smallest. For ease of notation in the following, suppose that the streams are re-indexed so that the sorted list of streams is $\tau_1,\tau_2,\ldots,\tau_n$. If $s_{\tau_1}\times(1- R_{\max})\leq W_0$ then there is enough unused budget to receive $\tau_1$ at full-bandwidth, so the stream $x_1$ has $s_{x_1}=s_{\tau_1}$ and contributes $q_1$ to the average quality. This leaves an unused bandwidth budget of $W_1=W_0-s_{\tau_1} \times (1- R_{\max})$ for the remaining streams after $x_1$. The algorithm repeats for $\tau_2, \tau_3,\ldots$ until some streams $\tau_{\ell}$ cannot be received at full bandwidth within the remaining budget $W_{\ell-1}$. It then determines the maximum bandwidth at which it can be received by calculating the minimum bandwidth reduction $r_{\ell}$ which similar to $R_{max}$ is a multiplicative inverse (reciprocal) of a power of $c$ such that $s_{\tau_{\ell}}\times r_{\ell}\leq W_{\ell-1}+ (s_{\tau_{\ell}}\times R_{\max})$. The received stream $x_{\ell}$ will have bandwidth $s_{x_{\ell}} = s_{\tau_{\ell}}\times r_{\ell}$ and will contribute $q'_{\ell}$ to the average quality of the system. The remaining bandwidth budget after streaming $x_{\ell}$ will be $W_{\ell} = W_{\ell-1} - (s_{\tau_{\ell}}\times r_{\ell}))$. The algorithm repeats this process to determine the bandwidth, amount of bandwidth budget, and quality contribution for each of the remaining streams $x_{\ell+1},x_{\ell+2},\ldots,x_{n}$. Finally the total quality and other statistics are calculated.

The other two heuristic algorithms (namely \textit{Aggressive} and \textit{Round-Robin}) are also represented in Algorithm 2 and Algorithm 3 respectively.
The algorithm needs a one-time implementation in the beginning of the session for the main process. Therefore it is implemented in real-time and does not provide any additional overhead during the runtime. It is implemented efficiently in $O(n log n)$ time and $O(n)$ space and produces solutions very close optimal. The approximation error depends on the difference between the bandwidth chosen for the first stream that cannot be received at full bandwidth and the remaining budget available to receive it. In our experiments, the approximation error never exceeded 1\%. Our proposed heuristic algorithms achieve global optimality, and can be generalized to the sending-based constrained scenarios as well. The adaptation model also supports dynamic cases for add/remove of cameras and users after the one-time implementation. Also, it should be noted that the adaptive scheme only changes the FPS, but not the quality/compression ratio of the 3D data which could more reduce the bandwidth requirement of streams. However, the our proposed optimization scheme could be adopted in similar way to controlling the compression ratio as well.

\section{Evaluation}
\label{sec:evaluation}
To evaluate our proposed adaptation algorithms, we developed TELEVIS, a visual simulation environment implemented in Java to showcase our view-aware prioritized framework. In specific, the objectives of development of TELEVIS are: a) to provide a visual simulation environment to model realistic interaction and collaboration among participants through view-aware prioritization in a large-scale interactive 3D tele-immersive system, b) to quantitatively evaluate the scalability of the proposed framework to ensure the fulfillment of real-time requirements of hundred-scale stream adaptations, and c) to evaluate the quantitative impact of our proposed view-aware framework in adapting to rapid varying bandwidth and distinguishing the quality of streams with different priorities. Prior to development of TELEVIS, tele-immersive researchers were using general-pur-pose modeling tools such as M\"{o}bius \cite{mobius}. M\"{o}bius is a tool for modeling the behavior of complex systems and to study the performance of networked systems, from telecommunication software to aeronautical and biological Systems. To the best of our knowledge, currently there is no single simulation environment that visually models the realistic interaction of participants, specifically targeting 3D tele-immersive systems.

Purely from a tele-immersive context, some activities are more collaborative compared to others. For example, in tele-immersive dancing or collaborative gaming, it is expected that the users are interacting with each other formed into dispersed groups of two or more, while in a tele-presence meeting, it is more probable that the users are facing eachother. In order to enable such realistic showcases, we modeled the interaction of participants through two sets of algorithms, \textit{facing} and \textit{clustering} algorithms, as a sketch of collaboration within a shared virtual space. As for facing, three different facing situations are taken into account: The user can face towards centroid of other people, he can face at least one other participant, or may have random facing direction. The view-aware prioritization is implemented as described in Section \ref{sec:methodology}, by which the prioritized participants are distinguished in different colors within TELEVIS. As another dimension, as various activities differ significantly in their notion of what constitutes an interaction and how to efficiently model them, we also adopted cluster analysis, to automatically classify participants to achieve the desired properties of interaction. We utilized Gaussian Mixture Models about centroid to describe the distribution of people by different density functions, through the following steps:
\begin{enumerate}
\item Taking $K$ different seeds (considered as means of our Gaussian distributions)
\item Estimation step: To calculate responsibility of each Gaussian for each participant
\item Maximization step: Using responsibility as weights, and move the mean of the centroid towards the weighted average of other participants
\item Repeat 2 and 3 until the Gaussians no longer move.
\end{enumerate}

Similar to our Gaussian Clustering, we also defined clustering in pairs in which paired participants are form-ed in closer distances, as well as defining uniform distribution to simulate uniformly distributed people inside the virtual space.

\begin{figure}[!t]
\vspace{2mm}
\centering
\includegraphics[width=\columnwidth]{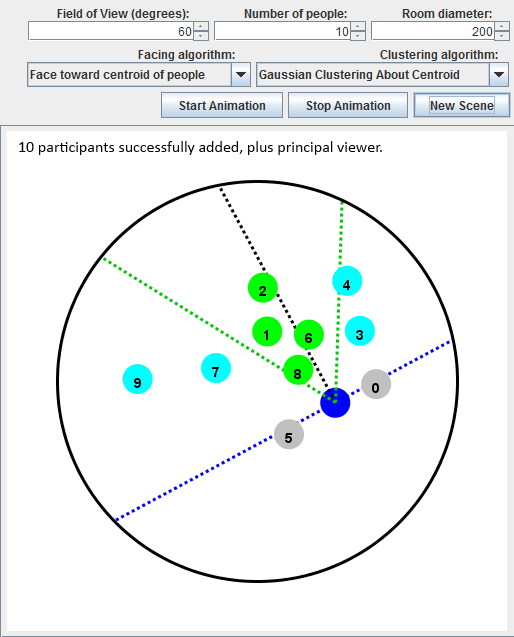}
\caption{Visual view of an example virtual space simulation in TELEVIS.}
\label{televis}
\end{figure}

\begin{figure}[!t]
\centering
\includegraphics[width=\columnwidth, trim = 50 225 50 225, clip = true]{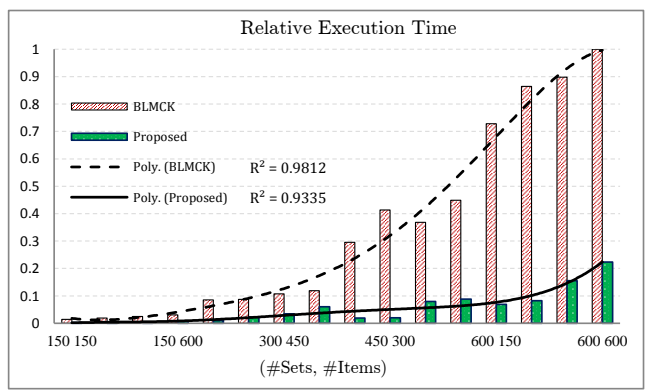}
\caption{A comparison of relative execution time needed for running BLMCK, and our proposed \textit{Compromise} algorithm to approximate multiple-choice knapsack problem for different number of sets and items inside each set.}
\label{comparison}
\end{figure}

To simulate the movement of participants, we implemented enhancement algorithms over the primary Random Walk Mobility Model \cite{probability}, \cite{sharma}. Let ${\{Z_k}\}_0^\infty$ be a sequence of independent, identically distributed discrete random variables. We implemented sequences $S_n=\sum {\{Z_i\}}^n$ as Random Walks of $Z_k$'s in $R^2$, corresponding to 2D area by definition. We view the sequence of $Z_k$'s as being the outcomes of independent experiments. Since the $Z_k$'s are independent, the probability of any particular (finite) sequence of outcomes can be obtained by multiplying the probabilities that each $Z_k$ takes on the specified value in the sequence. These individual probabilities are given by the common distribution of the $Z_k$'s in regards to three different modes, \textit{stay}, \textit{walk}, and \textit{turn} modes, which combined with our facing algorithm, provides intelligent movements to enable realistic modeling.

\begin{figure*}[!p!t]
\centering
\includegraphics[totalheight=0.21\textheight, trim = 45 240 45 240, clip = true]{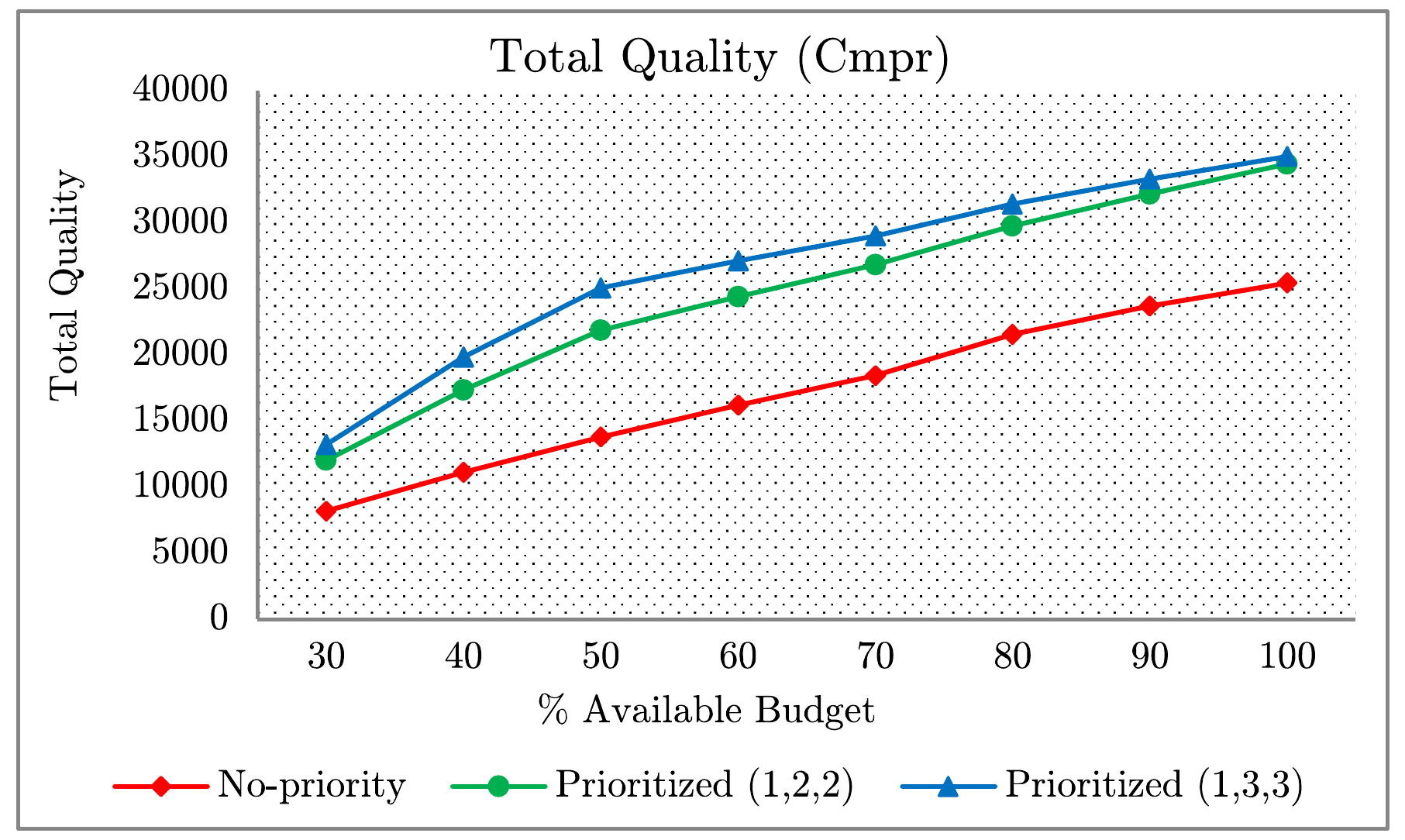}\includegraphics[totalheight=0.21\textheight, trim = 45 240 45 240, clip = true]{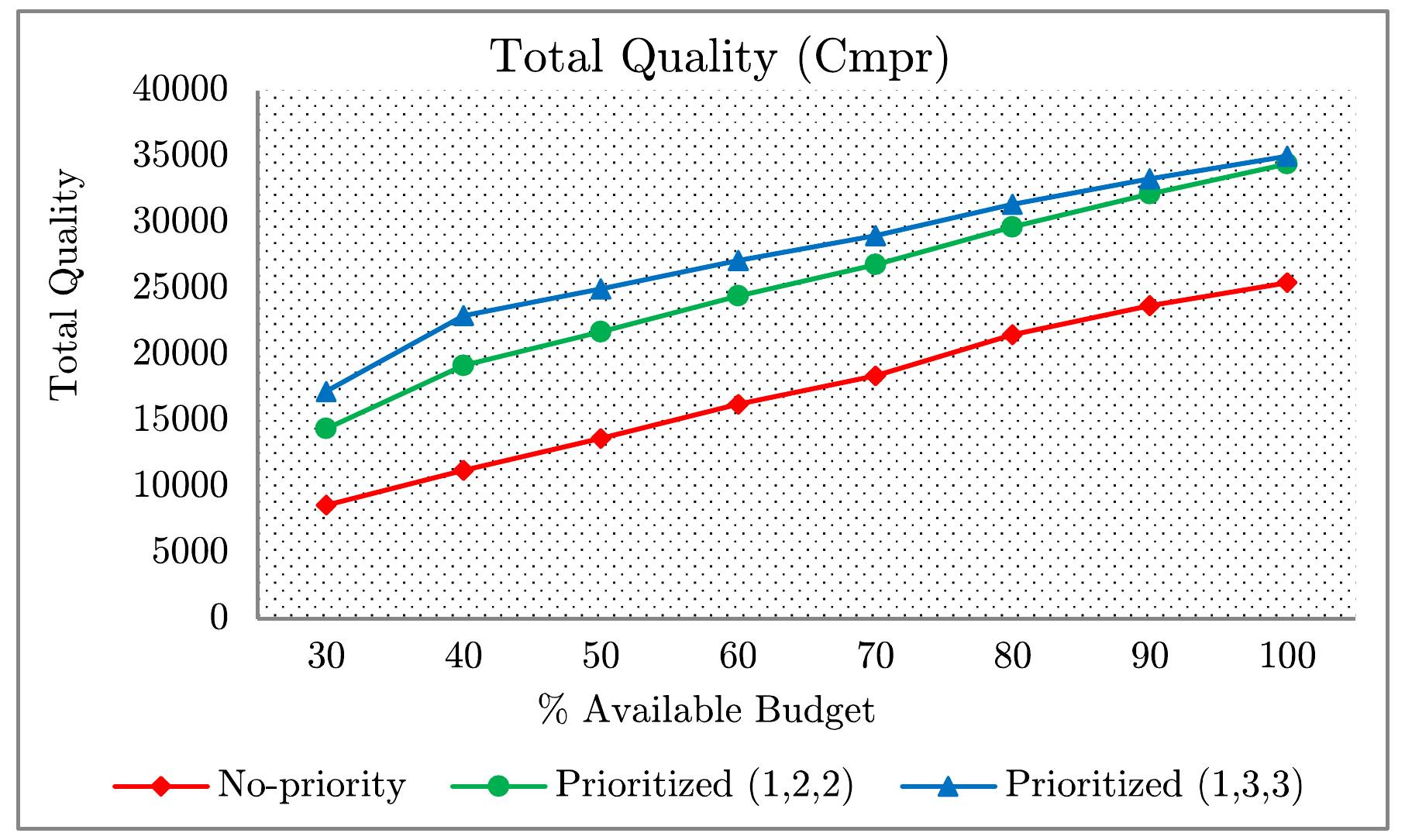}
\includegraphics[totalheight=0.21\textheight, trim = 45 240 45 240, clip = true]{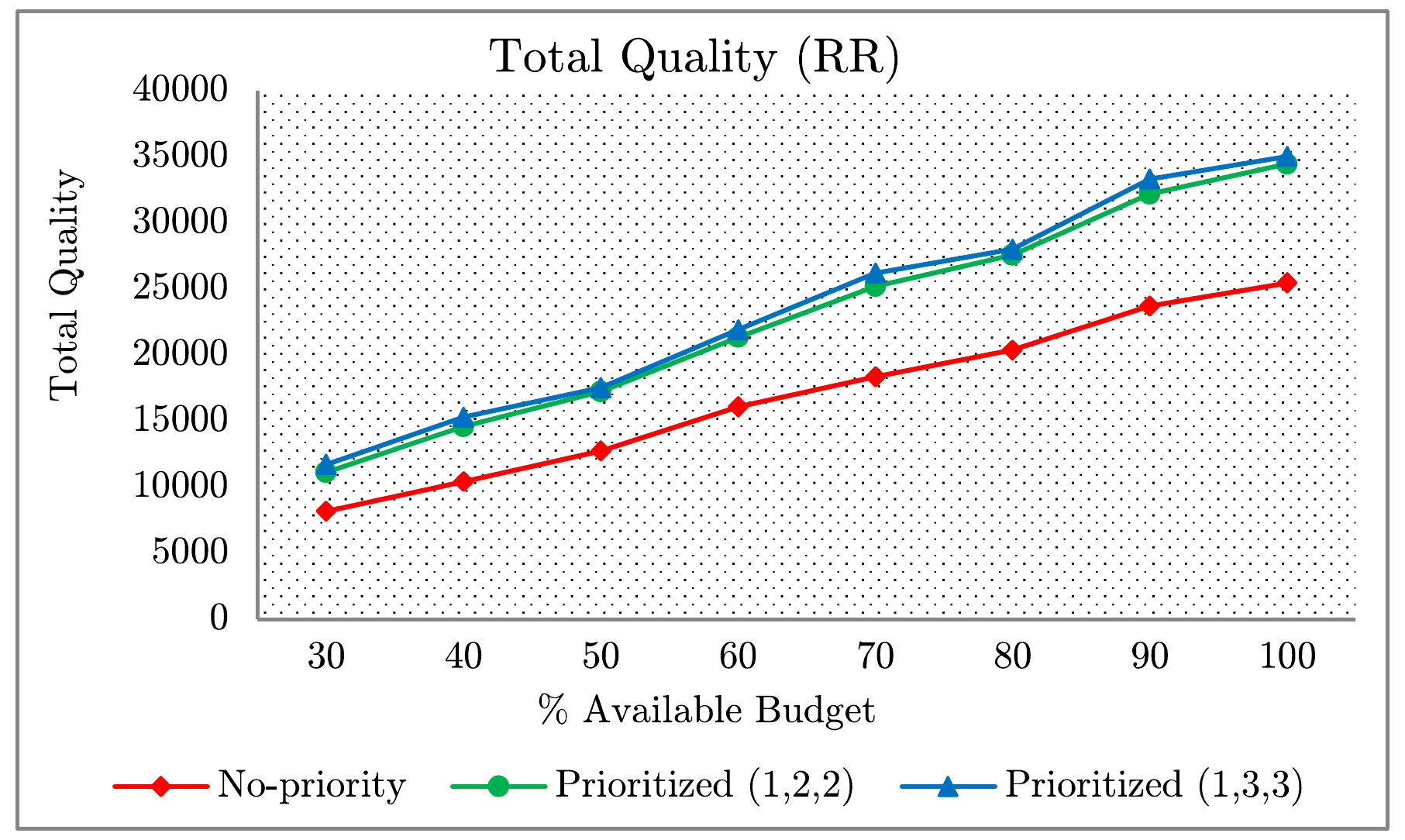}\includegraphics[totalheight=0.21\textheight, trim = 45 240 45 240, clip = true]{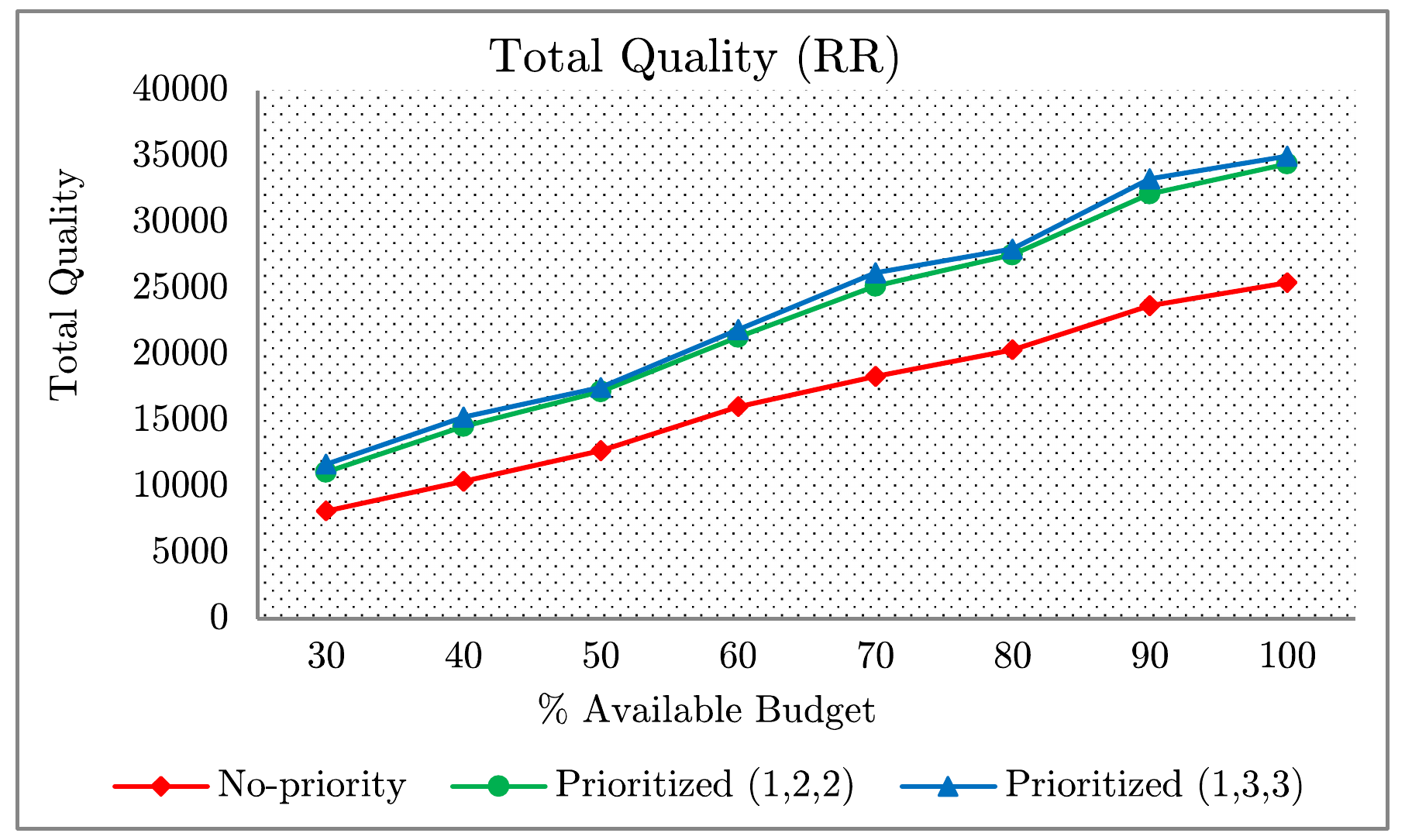}
\includegraphics[totalheight=0.21\textheight, trim = 45 240 45 240, clip = true]{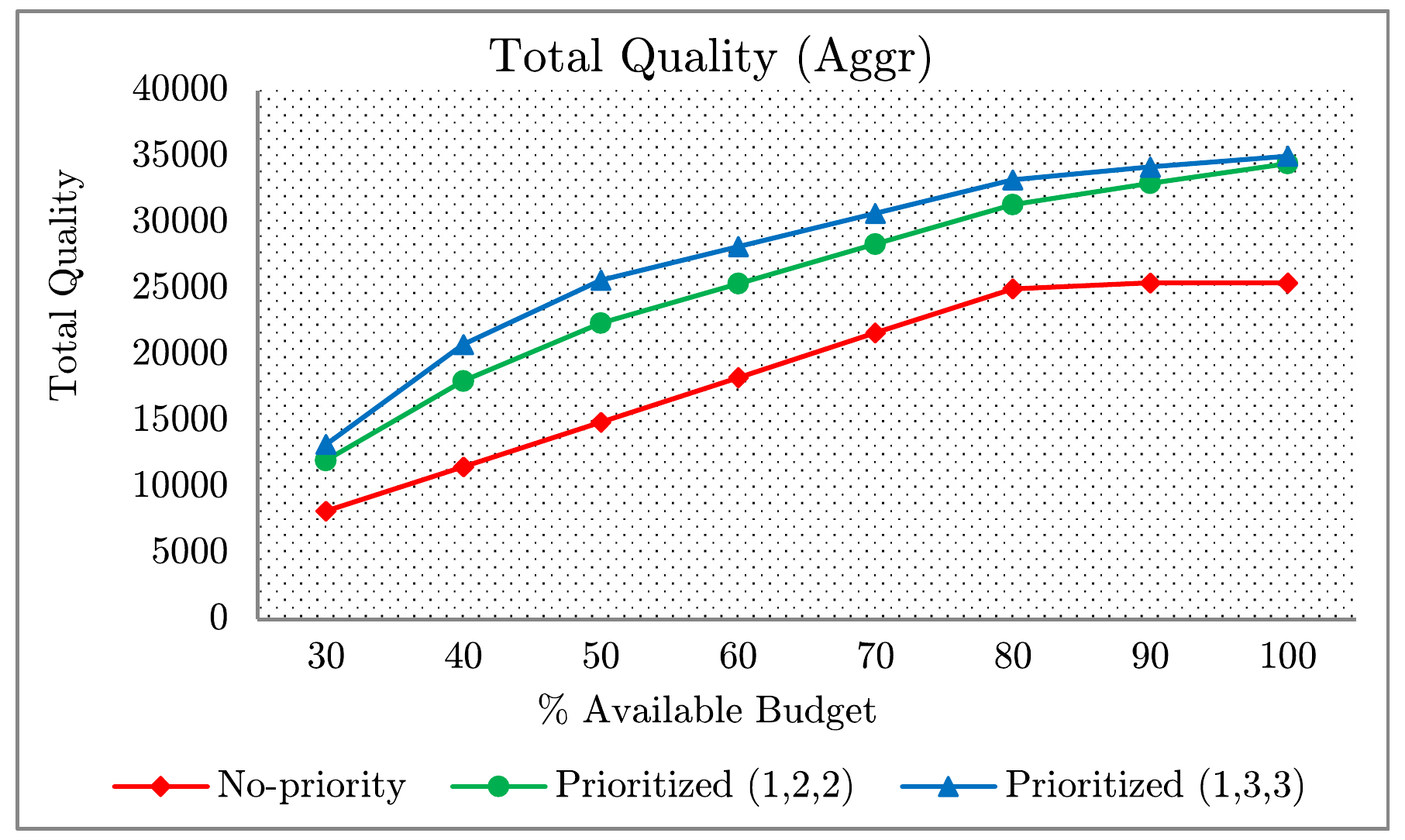}\includegraphics[totalheight=0.21\textheight, trim = 45 240 45 240, clip = true]{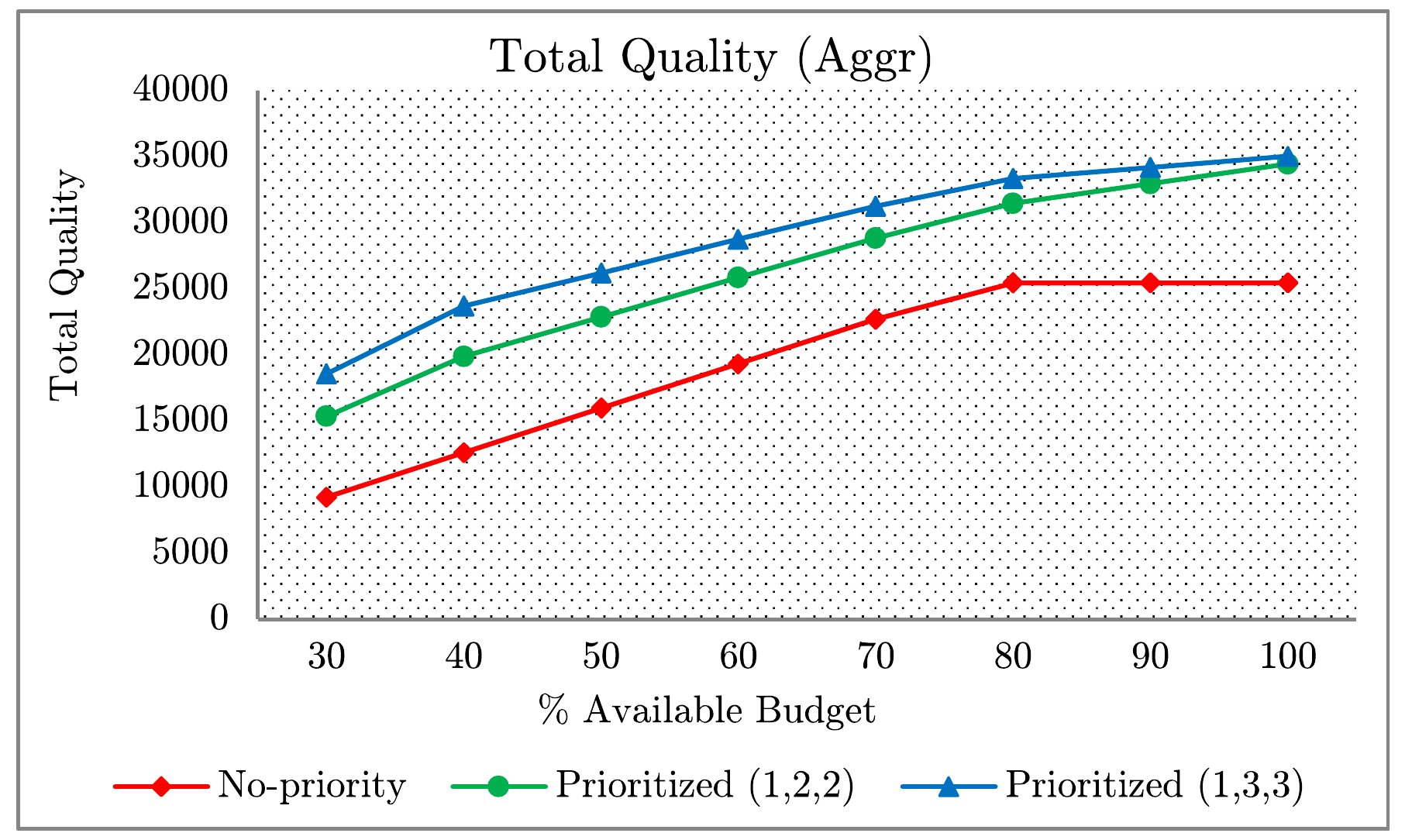}
\caption{Total quality measured for Aggressive, Round-Robin, and Compromise algorithms, for four priority classes and two values of $R_{max}=1/c^k$. (Top-Left) Compromise, $k=4$, (Top-Right) Compromise, $k=5$, (Middle-Left) Round-Robin, $k=4$, (Middle-Right) Round-Robin, $k=5$, (Bottom-Left) Aggressive, $k=4$, (Bottom-Right) Aggressive, $k=5$.}
\label{pdf:quality}
\end{figure*}

\begin{figure*}[!p!t]
\centering
\includegraphics[totalheight=0.21\textheight, trim = 45 240 45 240, clip = true]{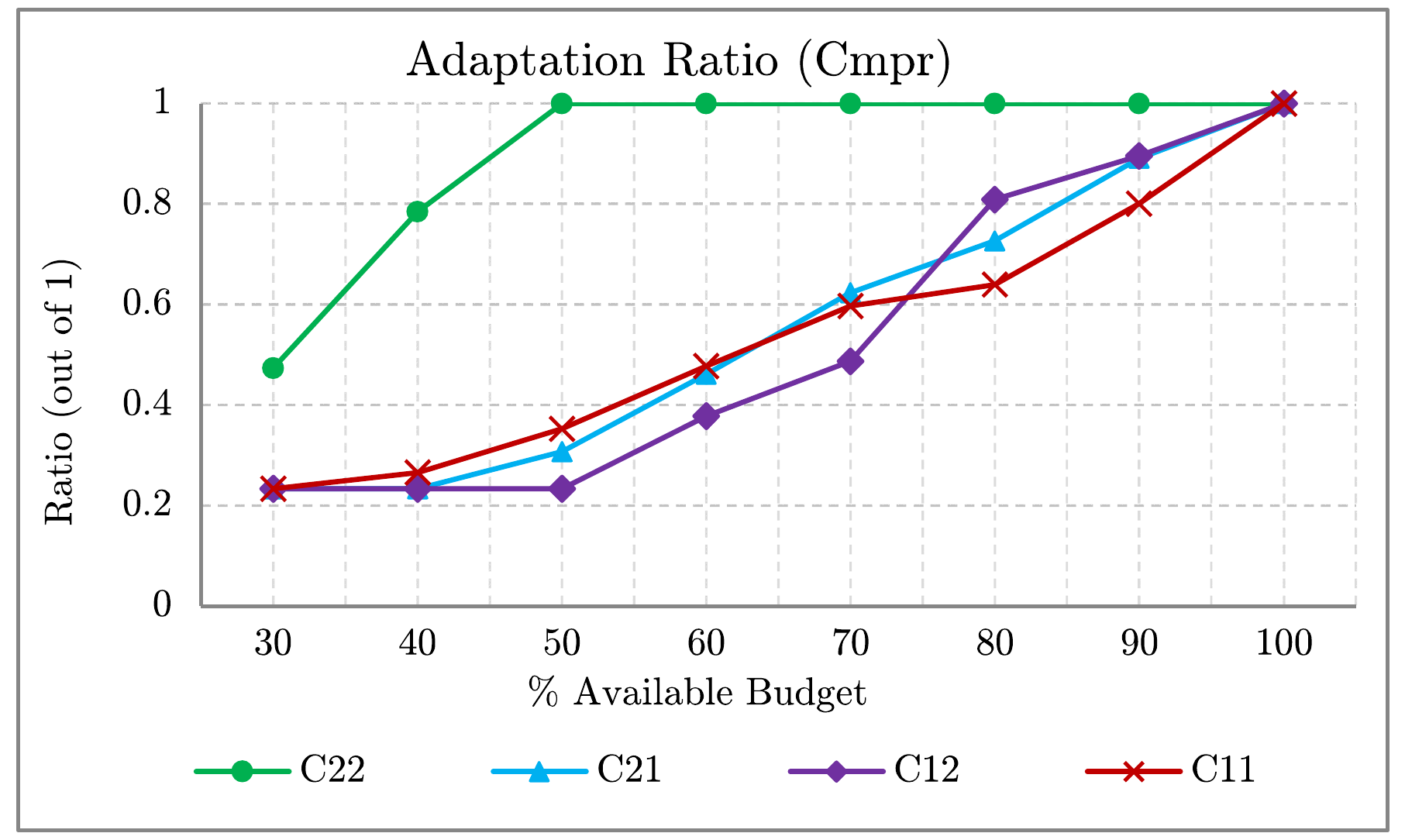}\includegraphics[totalheight=0.21\textheight, trim = 45 240 45 240, clip = true]{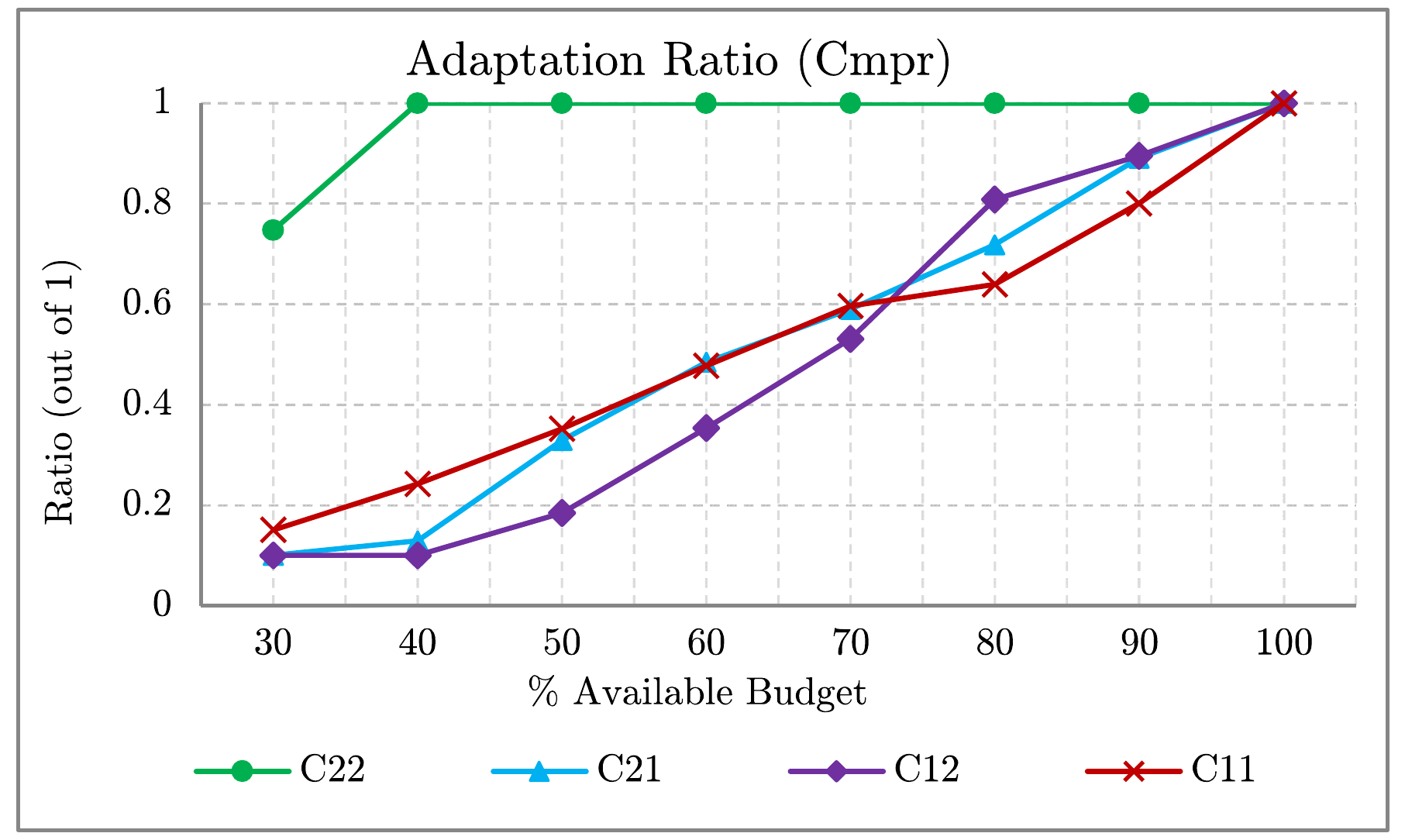}
\includegraphics[totalheight=0.21\textheight, trim = 45 240 45 240, clip = true]{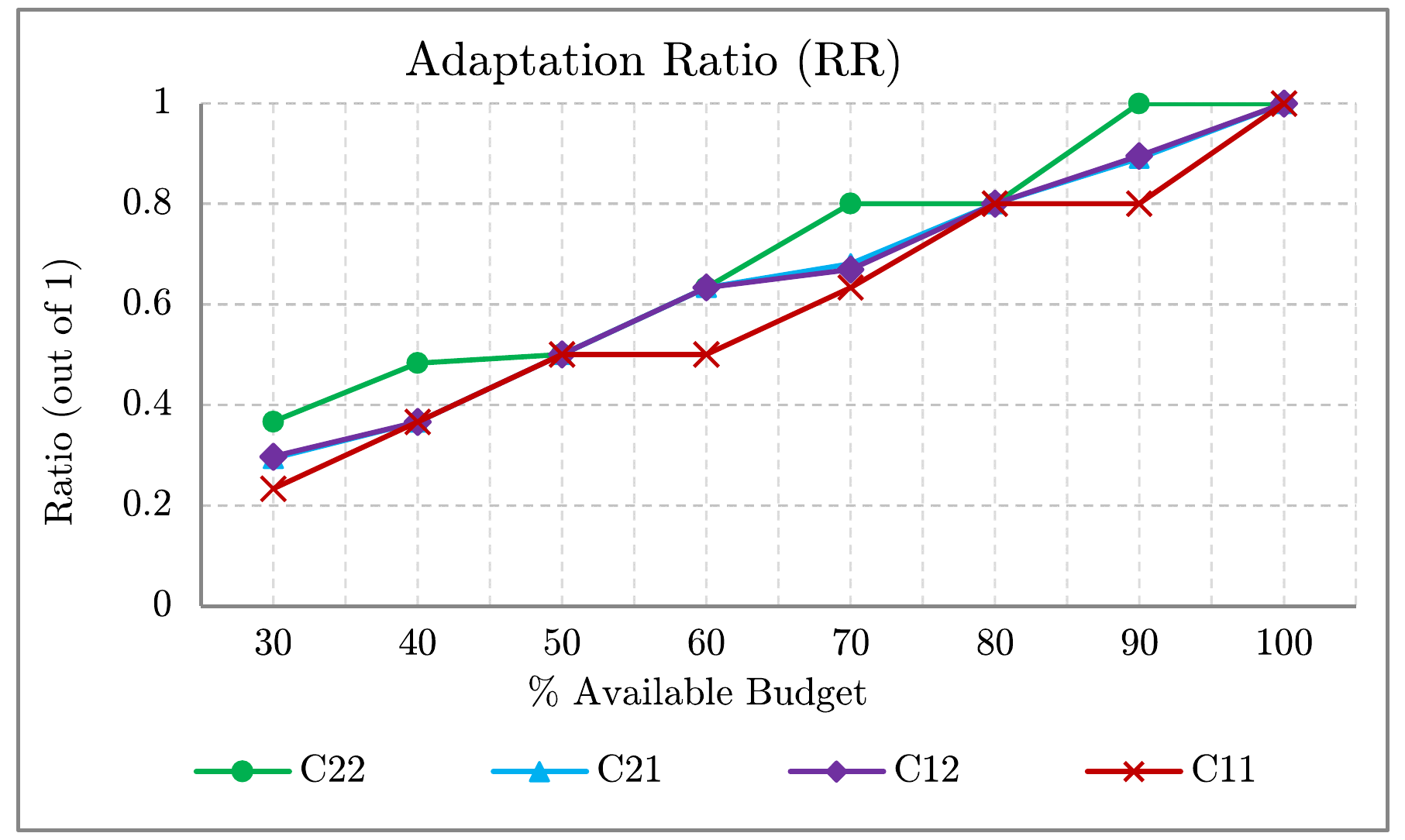}\includegraphics[totalheight=0.21\textheight, trim = 45 240 45 240, clip = true]{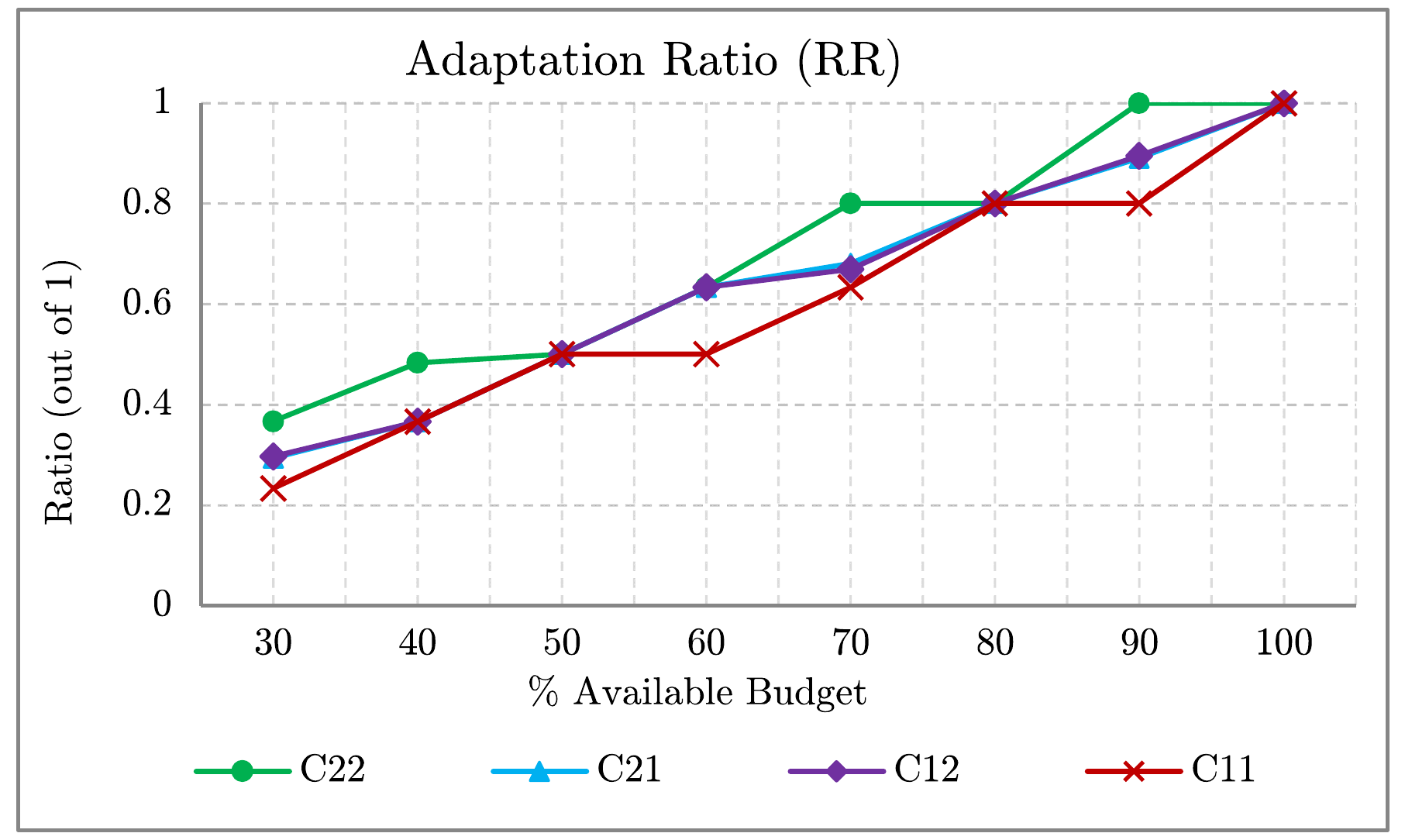}
\includegraphics[totalheight=0.21\textheight, trim = 45 240 45 240, clip = true]{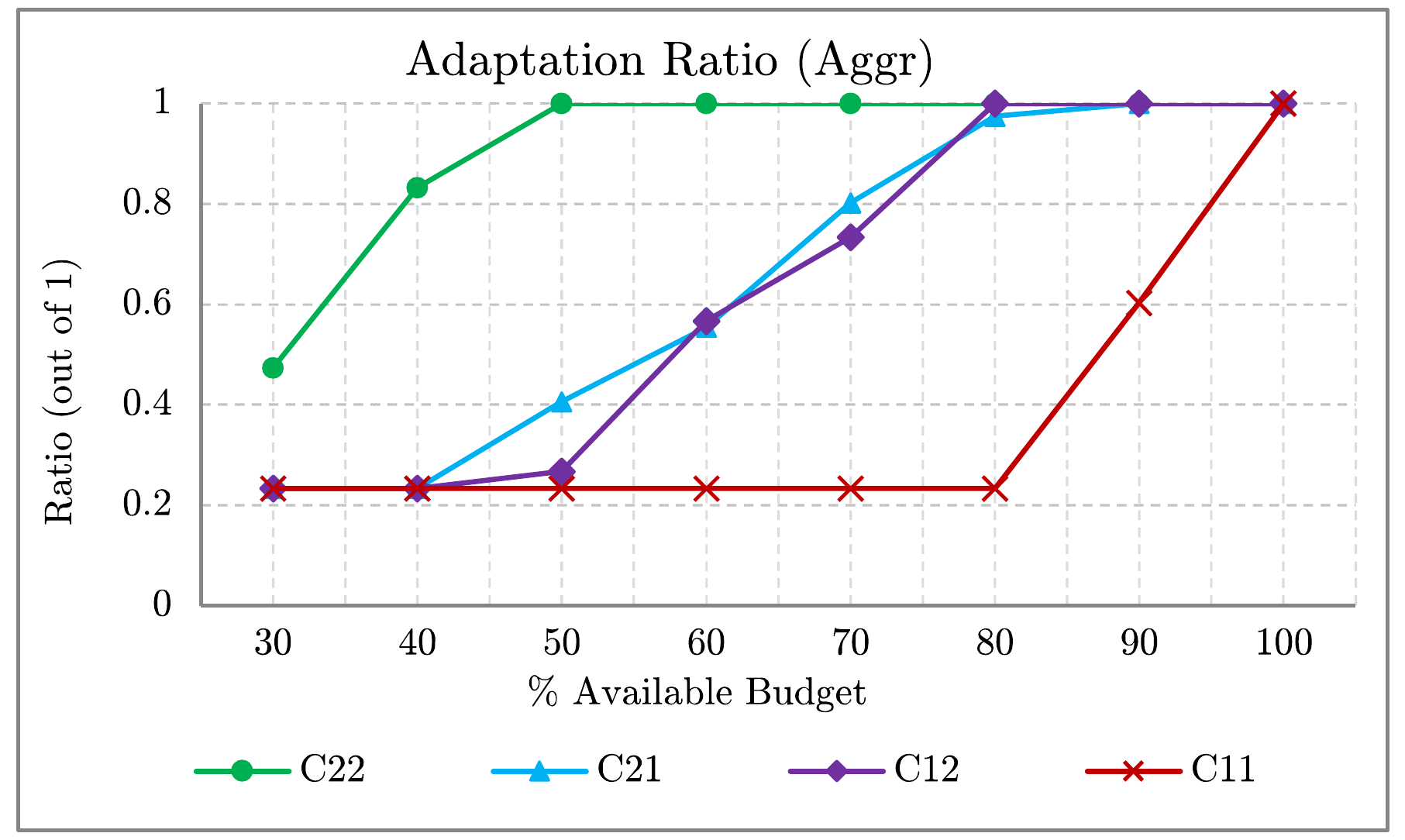}\includegraphics[totalheight=0.21\textheight, trim = 45 240 45 240, clip = true]{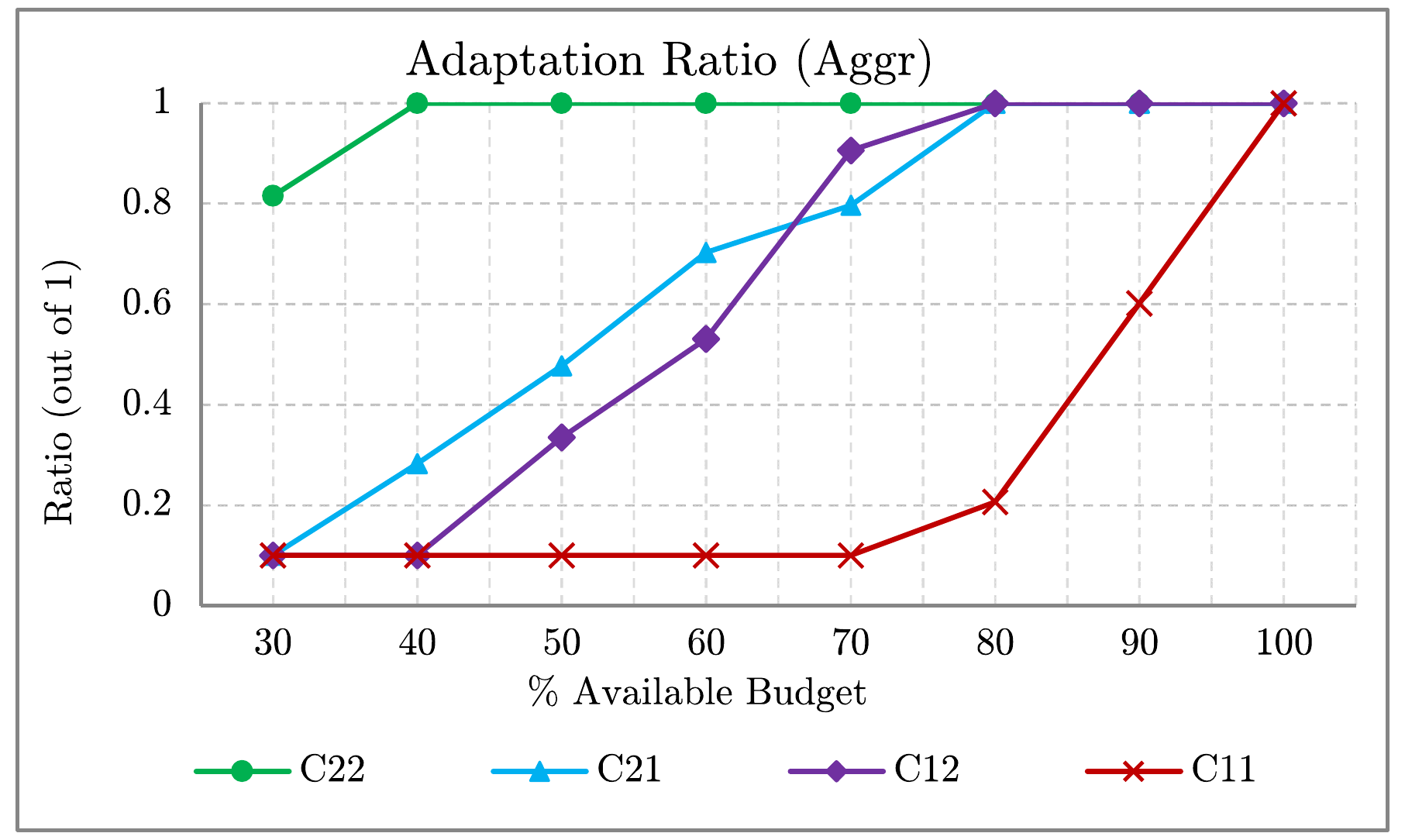}

\caption{Adaptation ratio in terms of bandwidth reduction, measured for four priority classes, and two values of $R_{max}=1/c^k$. (Top-Left) Compromise, $k=4$, (Top-Right) Compromise, $k=5$, (Middle-Left) Round-Robin, $k=4$, (Middle-Right) Round-Robin, $k=5$, (Bottom-Left) Aggressive, $k=4$, (Bottom-Right) Aggressive, $k=5$.}
\label{pdf:adaptations}
\end{figure*}

\begin{figure*}[!p!t]
\centering
\includegraphics[totalheight=0.21\textheight, trim = 45 240 45 240, clip = true]{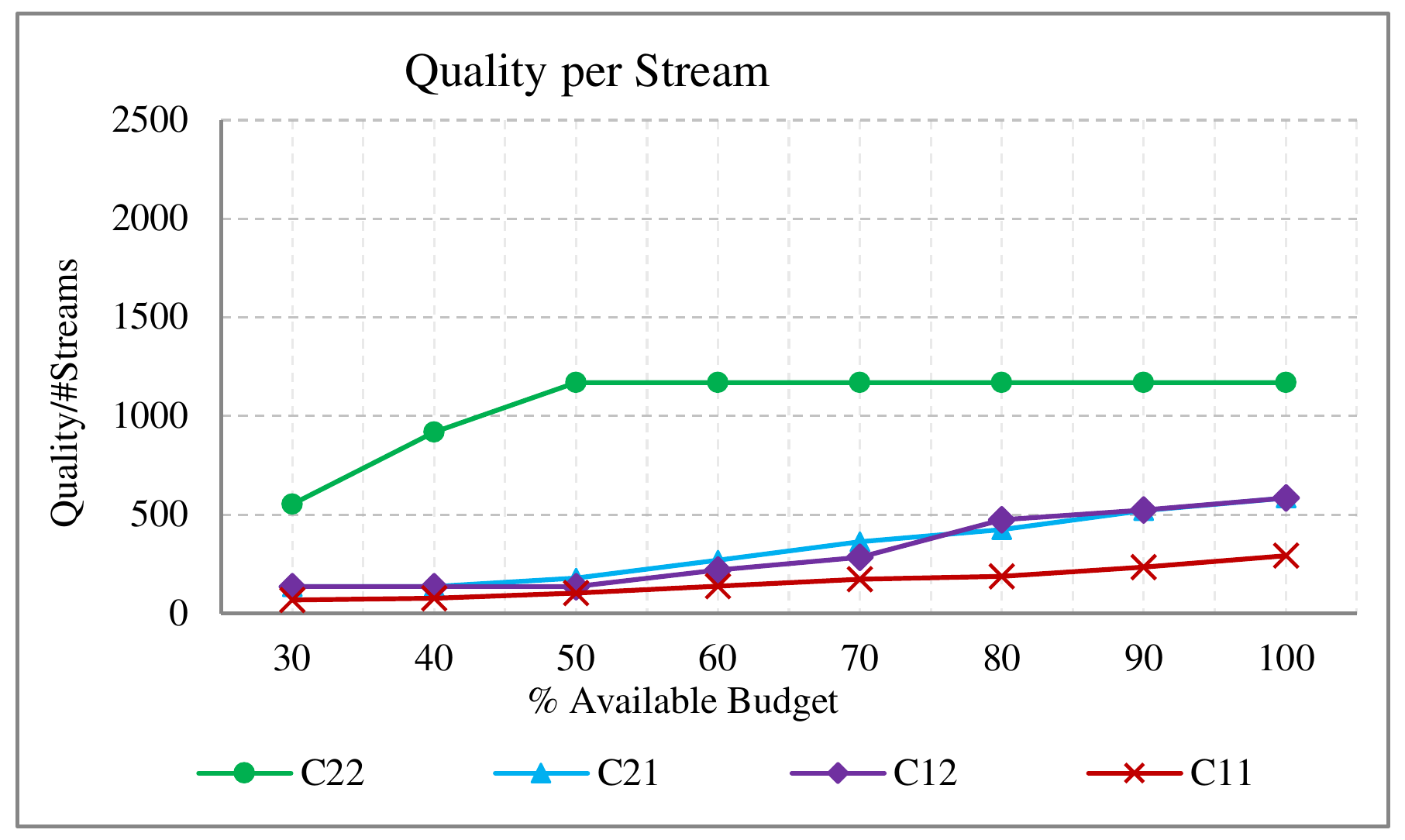}\includegraphics[totalheight=0.21\textheight, trim = 45 240 45 240, clip = true]{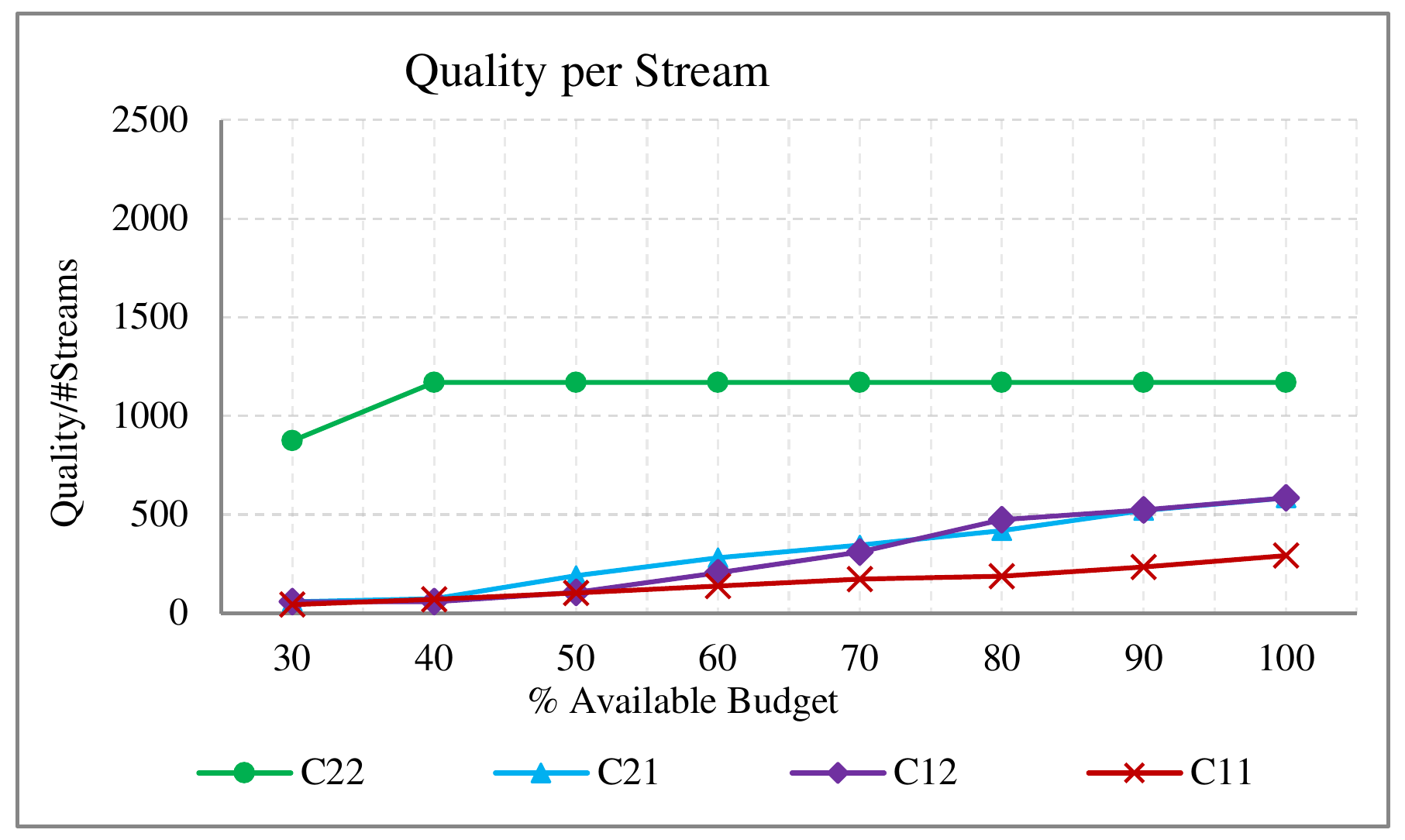}
\includegraphics[totalheight=0.21\textheight, trim = 45 240 45 240, clip = true]{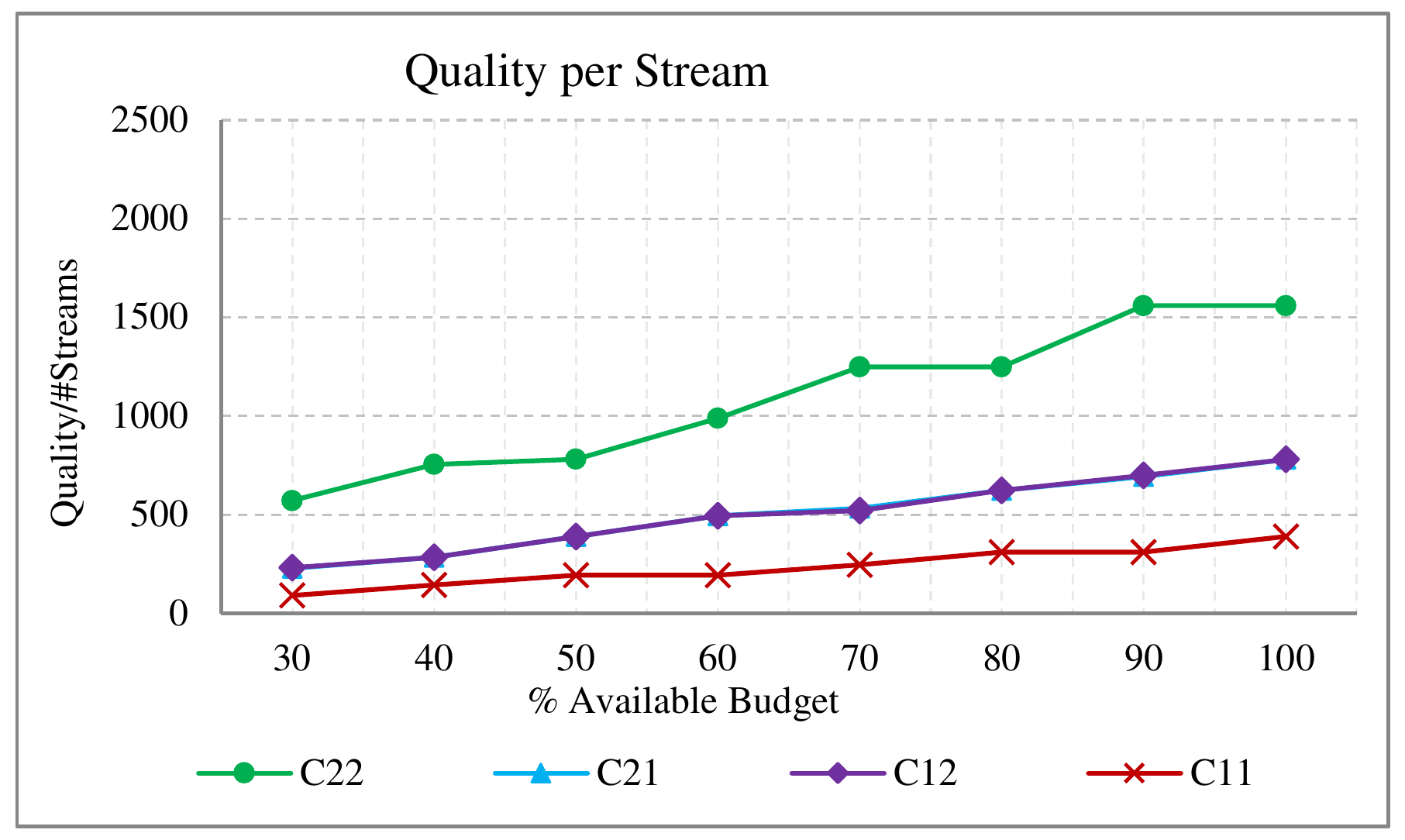}\includegraphics[totalheight=0.21\textheight, trim = 45 240 45 240, clip = true]{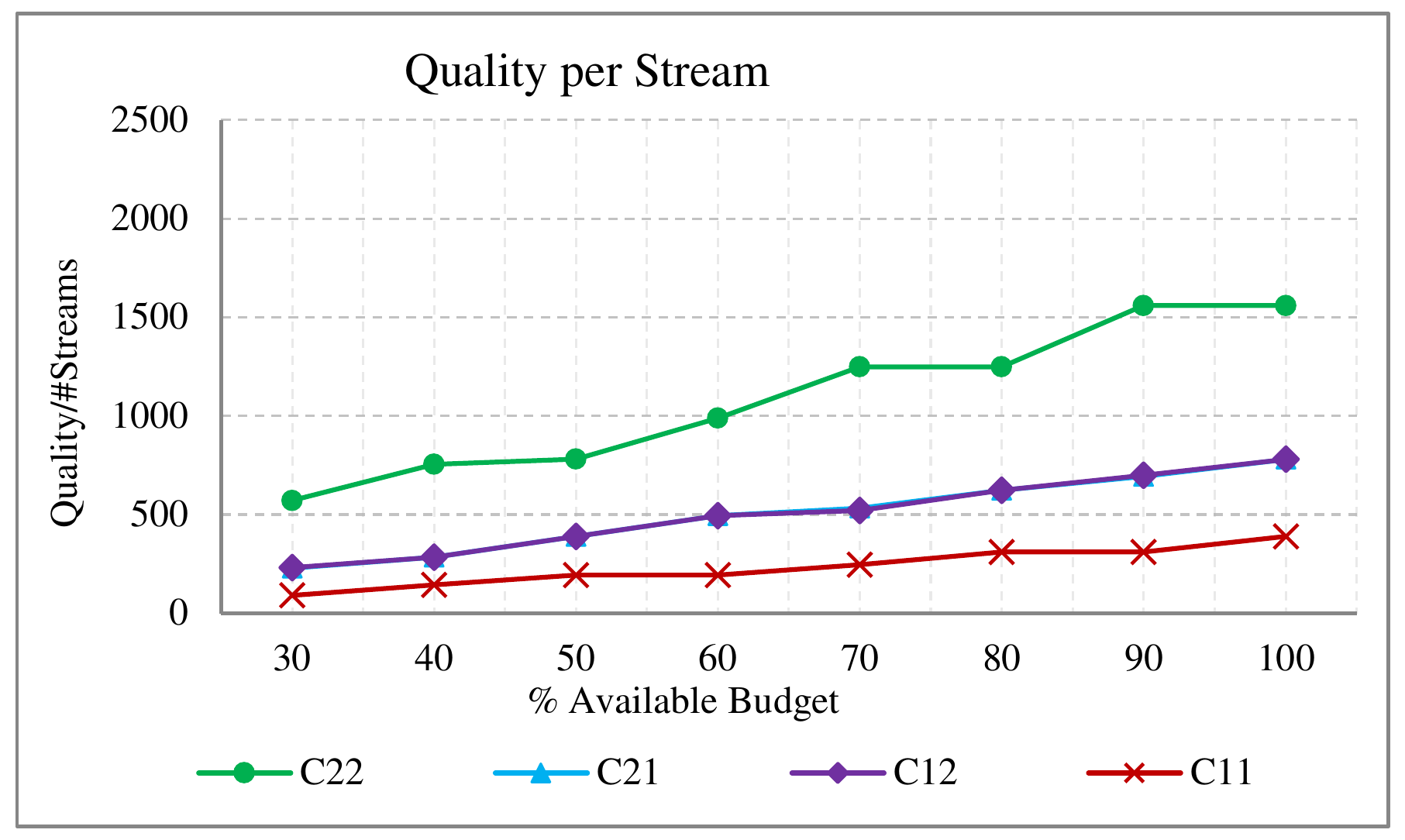}
\includegraphics[totalheight=0.21\textheight, trim = 45 240 45 240, clip = true]{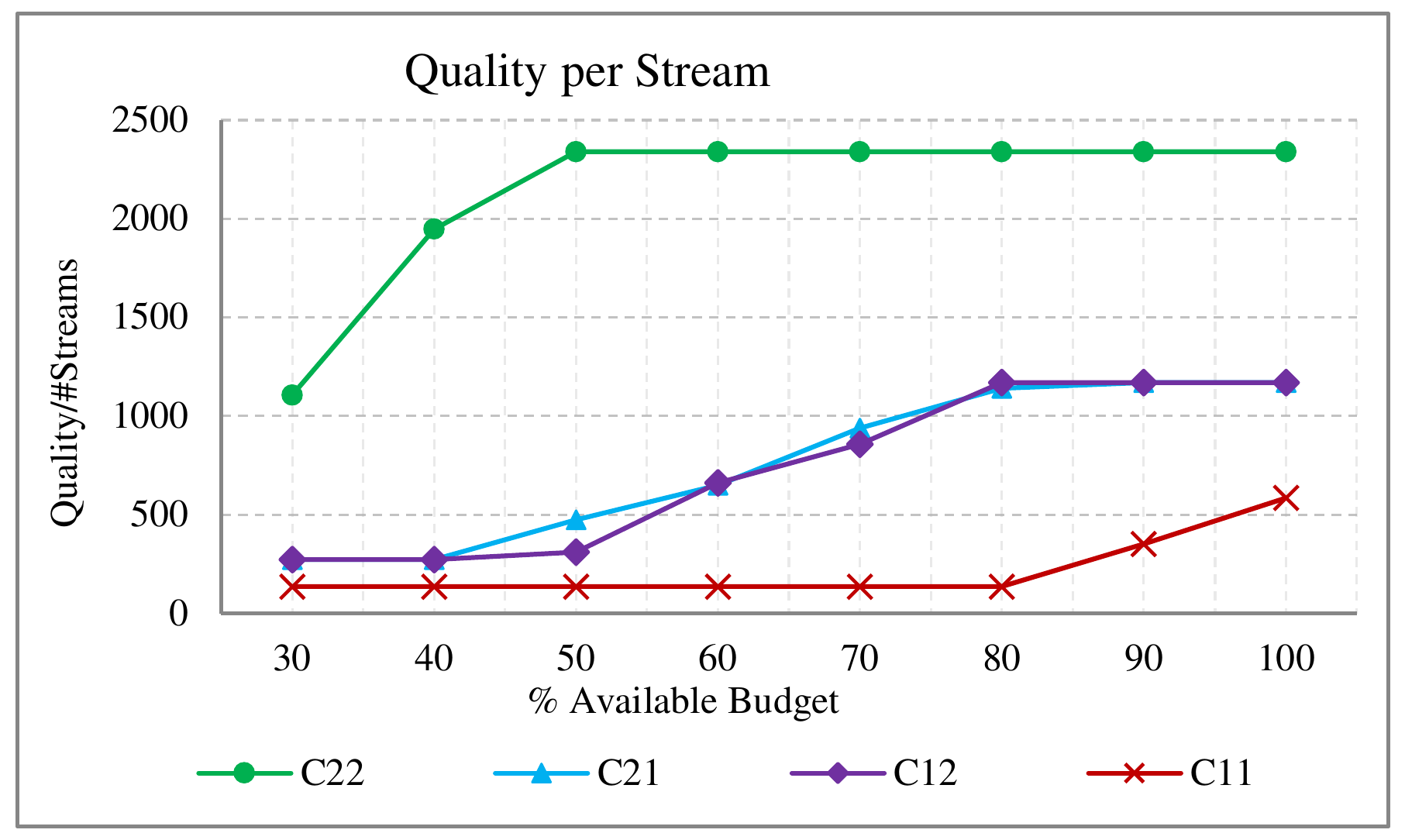}\includegraphics[totalheight=0.21\textheight, trim = 45 240 45 240, clip = true]{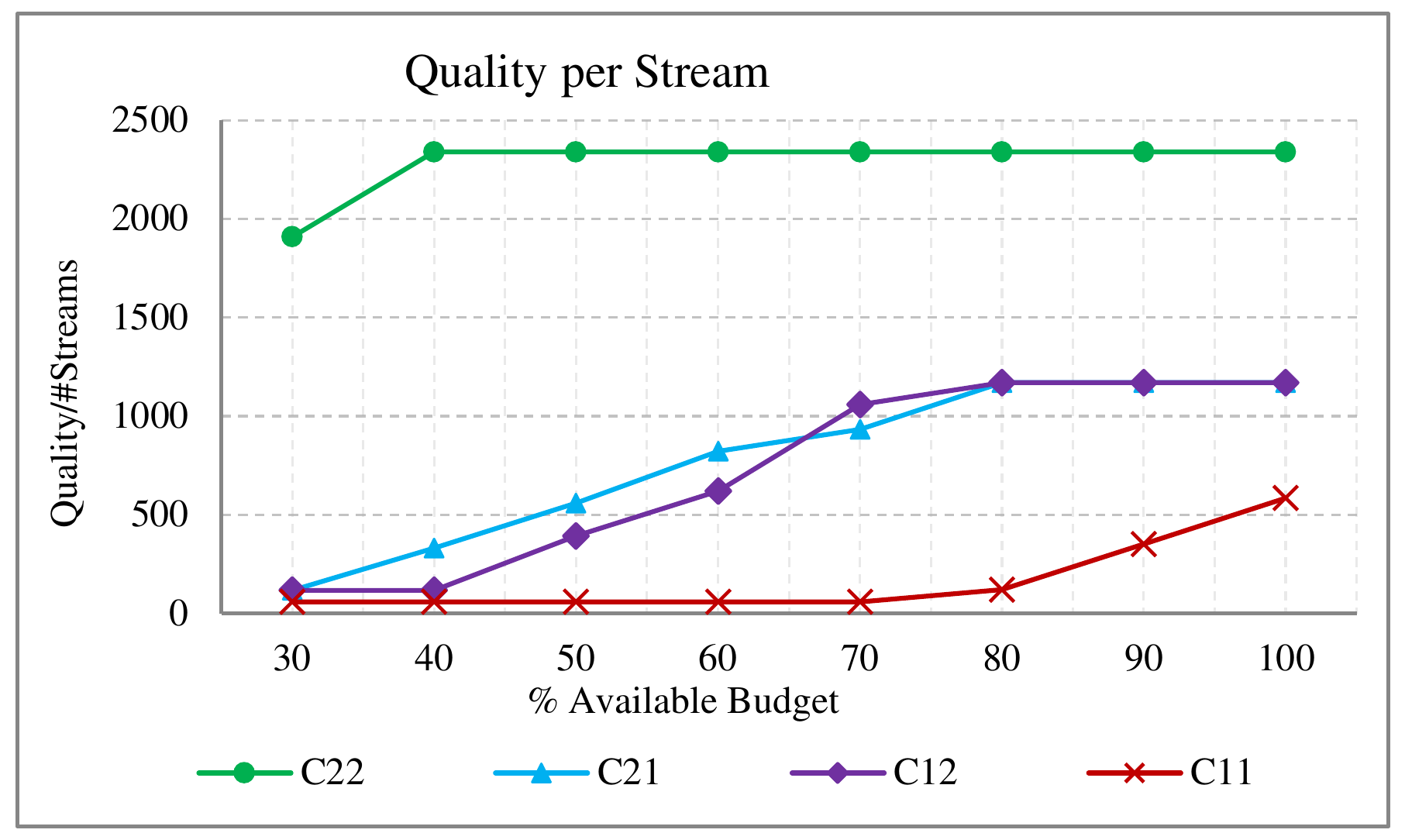}

\caption{Average quality per stream measured for four priority classes, and two values of $R_{max}=1/c^k$. (Top-Left) Compromise, $k=4$, (Top-Right) Compromise, $k=5$, (Middle-Left) Round-Robin, $k=4$, (Middle-Right) Round-Robin, $k=5$, (Bottom-Left) Aggressive, $k=4$, (Bottom-Right) Aggressive, $k=5$.}
\label{pdf:qps}
\end{figure*}

Figure \ref{televis} shows an example visual overview of a virtual space with 10 participants and their priorities in regards to the primary viewer's FOV as simulated in TELEVIS. This figure only shows the virtual viewpoint for the primary user (dark blue). Other users' viewpoint direction was not simulated here. TELEVIS enables management of various control parameters, such as the number of participants, FOV degrees, room diameter, bandwidth budget, etc. as well as different algorithms including the facing and clustering algorithms, which overall characterizes realistic interactions. It should be noted that it is assumed that the users not only traverse inside the virtual space, but can also move within their physical space in front of the cameras. 

We evaluated the execution runtime of our proposed \textit{Compromise} algorithm, and examined the average-case performance of the algorithm by analyzing the computational experiments that were conducted. Kozanidis \textit{et al.} \cite{blmck} explored the fundamental properties of multiple-choice knapsack problems, and presented a mathematical formulation to develop an efficient algorithm, termed BLMCK. We compare our main algorithm with BLMC-K. Figure \ref{comparison} shows a relative comparison of execution times for different sizes of multiple-choice sets and items inside each set, denoted by $(N, M_k)$, with the maximum execution runtime scaled to 1. Assume $N$ be the number of multiple choice sets (i.e. sites), and $M_k$ be the number of items in multiple choice set $k$ (i.e. cameras). In these experiments, $M_k$ was assumed to be the same for all $k$. Both $N$ and and $M_k$ varied between 150 and 600 in steps of 150. This means that the biggest problem solved contained 360,000 decision streams. The curve lines show the polynomial trend-line for the relative execution time, indicating how accurate the trend-lines are in generalizing the overall trend and future data-point estimation. As can be seen, the results illustrate the efficiency of our proposed \textit{Compromise} algorithm, which outperforms BLMCK, and makes it more suitable for the real-time requirement of stream adaptation.

To evaluate our proposed adaptation algorithms, we used 10 experimental tests derived from 10 participants (i.e. sites) and 10 maximum number of cameras within each site, and calculated performance measures, and the mean and standard deviation of the results. Our performance measure is total quality, as calculated in Section \ref{sec:methodology}, which is a measure of the effectiveness of an approach to maximizing the total amount of data based on prioritizations, with larger values being more effective. We ran our experiments with the available budget $W$ set to be different percentages of \textit{S} (total bandwidth of all streams). In particular, we set $W$ to 0.3\textit{S}, 0.4\textit{S}, ..., 1.0\textit{S} corresponding to 30\%, 40\%, ..., 100\% of the total bandwidth of streams. We set $R_{max}=\frac{1}{\ceil*{c^k}}$ in which we chose $c=\sqrt{2}$ based on a trade-off between the different choices of FPS and a decreasing fine-grained gaps between them, and tested two values for $R_{max}$. In practice, the 3D tele-immersive viewer will choose $R_{max}$ based on the perceived quality of the streams. We also determined the contributions to total quality of the streams in each of the four priority class $C_{11}$, $C_{12}$, $C_{21}$, and $C_{22}$ for all three algorithms. We repeated each trial of our simulation experiments for several times to ensure that the standard deviation of the measurements was within acceptable limits. Figures \ref{pdf:quality} to \ref{pdf:qps} show a small sample of our experimental results. Our other experimental results show similar trends.

Figure \ref{pdf:quality} shows the experimental results for the total quality achieved using our three adaptation algorithms. 
We used two different triples ($p_0$, $p_1$, $p_2$) of first-level and second-level priorities relative to $p_0$, (1,2,2) and (1,3,3), to differentiate the priorities of streams in different classes. We set $p_0=1$ to normalize the priorities. As can be seen, the total quality achieved by all three algorithms increases as the available budget increases because bandwidth is a component of our definition of quality. As the available bandwidth budget increases, there is more space for larger stream bandwidth. Also, the total quality increases as the ratio of $\frac{p_1}{p_0}$ or $\frac{p_2}{p_0}$ increases, confirming that our proposed prioritized approach noticeably distinguishes the important streams from the less important ones. Interestingly, it can be concluded that ratios between the smallest and largest priorities that are larger than 1:3 are not likely to be interesting. Also, in the graphs for \textit{Aggressive} and \textit{Compromise}, the maximum gain in total quality is in the range of 30\% to 50\% of the available budget, suggesting that this the range where our approach is most effective.

Figure \ref{pdf:adaptations} shows the results for average bandwidth reduction in terms of how much the adaptations scaled down the bandwidth, given the equation

\begin{equation*}
Adaptation~Ratio: \dfrac{\text{total quality \textit{after} adaptation}}{\text{total quality \textit{before} adaptation}}
\end{equation*}

measured for four different pairs indicating each priority class, and two different values of $R_{max}=1/c^k$ ($k=4$ and $k=5$). As can be seen, the larger value of $R_{max}$ results in more bandwidth reduction, and thus quality sacrifices, for streams in $C_{22}$ (corresponding to the high-priority class), while preserving more of the full-bandwidth streams in $C_{11}$ (corresponding to the low-priority class). Clearly the larger value of $R_{max}$ works better for lower values of the available bandwidth. Also, similar to the total quality, these figures suggest two points. Firstly, the adaptation ratio increases as the available budget increases. Secondly, the diagram of $C_{11}$ shows larger adaptation ratio compared to $C_{22}$, confirming that our proposed approach noticeably distinguishes the important streams from the less important ones, with \textit{Aggressive} and \textit{Round-Robin} brin-ging the most, and the least differentiation, respectively, as compared to the compromise-based differentiation given by \textit{Compromise}. 

Figure \ref{pdf:qps} shows average quality per each stream for the same set of experiments. The diagrams confirm the results derived from Figure \ref{pdf:quality} and Figure \ref{pdf:adaptations}. As can be seen, the diagram of $C_{22}$ shows larger values compared to the other priority classes. It is worth mentioning that our approach carries maximum effectiveness when all the streams in $C_{11}$ are adapted by $R_{max}$, and none of the $C_{22}$ streams are adapted. This specific point is considered as the peak of quality. As our framework adapts the streams in $C_{22}$, the gain in quality brought by our approach is being decreased.

However, our framework does provide quality degradation in general, but considering the bandwidth savings achieved using our adaptations, it is reasonable to believe that tele-immersive users would make this sacrifice in quality in exchange for respecting their view, scalability, and the network bandwidth constraints.

\section{Conclusion}
\label{sec:conclusion}

In this paper we studied an adaptive prioritized HVS-compliant framework to manage delivery of hundred-scale 3D tele-immersive streams to a receiver site with limited bandwidth constraint. Our adaptation framework exploits the semantics link of FOV with multiple 3D streams in a 3D tele-immersive environment. We developed TELEVIS, a configurable visual simulator to showcase a HVS-aware tele-immersive system for realistic cases. The evaluation results show that our adaptations can improve the total quality of streams in 3D tele-immersive systems by making best use of the limited bandwidth.

In the future, we plan to derive a more sophisticated and realistic measures of quality, and refine our algorithms by conducting more extensive experiments, including experiments on a real 3D tele-immersion testbed to measure the actual bandwidth limits and bandwidth usage of streams. We would like to rigorously verify the worst-case performance of our algorithms using run-time verification techniques~\cite{ahmadyan}. We also plan to add transmission time into the formulation of our adaptations to be featured in TELEVIS. We would also like to apply our design directly to a general display as opposed to just HMDs, and take into account the dependency of FOV on direction of the gaze and distance from the screen.

\begin{acknowledgements}
This research is funded in part by NSF CNS 13-29886 and in part by Navy N00014-12-1-0046.
\end{acknowledgements}

\bibliographystyle{spmpsci}
\bibliography{sigproc-Revised}

\end{document}